\documentclass[12pt,epsfig]{article}
\usepackage{epsfig}
\newcommand{\s}{$\bar{s}s$}
\newcommand{\p}{$\bar{p}p$}
\newcommand{\f}{$\phi$}
\newcommand{\an}{annihilation}
\newcommand{\ap}{antiproton}

\newcommand{\om}{$\omega$}
\newcommand{\ten}{$f'_2 (1525) $}
\newcommand{\la}{$\Lambda$}
\newcommand{\al}{$\bar{\Lambda}$}

\begin{document}

\begin{center}

{\Large\bf{

Experimental tests of the Okubo-Zweig-Iizuka rule
in hadron interactions}}

\vskip1cm

V.P.Nomokonov, M.G.Sapozhnikov\\~\\

Laboratory of Particle Physics,\\
Joint Institute for Nuclear Research, Dubna

\end{center}

\vskip2cm

\section{Introduction}

        The Okubo-Zweig-Iizuka rule (OZI) \cite{OZI} was suggested at the early
stage of the development of the quark model of hadrons. G.Zweig would like to
explain why the $\phi(1020)$ meson, which
mass is barely above the mass of $K\bar{K}$ system, decays preferentially into
two kaons but not into $\pi \rho$ channel in spite of larger phase space.
He proposed  that \f~ has an internal structure which is more similar
to the kaon  than to the pion and assumed that the strange quarks
of $\phi$ prefer to maintain their identity rather than transform to light
quarks.

        S.Okubo described this tendency of light quarks to keep their identity
as a rule which forbids creation of \s~
mesons in the interactions of baryons and mesons consisting of
$u$ and $d$ quarks only.

        Nowadays a popular formulation of the OZI rule is that
the reactions with disconnected quark lines are suppressed.
There are also other formulations of the rule, which are nicely reviewed
in \cite{Lip.91}.

        Since its appearance the consequences of the OZI rule  have been
 tested in a number of
experiments. Different probes were used in a wide interval of energies.
It was found that in practically all of these reactions the
OZI
rule is fulfilled well, within  few percent accuracy.

        The question arises: why? The OZI rule was suggested as a purely phenomenological
rule. The reasons why it works well was not understood till the
development of the QCD.
Though we have not still a full understanding of
the success of the OZI rule, it is clear now why it should
be fullfilled in some limits. For instance, in the limit of heavy quarks
and in the limit of large number of colours $N_c$.
In these cases the OZI rule appears automatically from the QCD general
principles.

        Now the OZI rule is well tested in many experiments and
has solid theoretical background. In this situation it is quite intriguing
the results of experiments with stopped antiprotons at LEAR (CERN), where
unexpected large violation of the OZI rule was found (for the review,
see \cite{Ams.98},\cite{Sap.98}).
In some annihilation channels
 the $\phi$-meson production exceeds the prediction of the OZI rule
by a factor of 30-70.
It is remarkable that the level of the OZI rule violation
 turns out to be different in various
 annihilation reactions.
For instance, in \ap~\an~ at rest the observed yield
of the reaction $\bar{p}p \to \phi \pi$
exceeds the OZI rule by a factor of 30,
but in the similar reactions  $\bar{p}p \to \phi \rho$ or
$\bar{p}p \to \phi \omega$
no violation has been found.

It turns out that the deviation from the OZI rule prediction
depends  drastically  on the quantum numbers of the initial state
 of the nucleon-antinucleon system.
 Thus, the reaction $\bar{p}p \to \phi\pi^0$ at rest
 is allowed from two $\bar{p}p$
 initial states, $^3S_1$ and $^1P_1$.
It was found a strong violation of the OZI rule from $^3S_1$ state,
 but for \an~
from  $^1P_1$~state the \f~ production is at least 15 times less,
which is in agreement with the OZI rule prediction.

        Strong violation of the OZI rule was observed not only in the
 \ap~\an, but also in the reactions with protons
 $pd \to~ ^3He \phi$
\cite{Wur.95}, $pp \to pp \phi$ \cite{Bal.98},\cite{Bal.00} and pions
$\pi p \to \phi \pi p$ \cite{Fer.97}. The discrepancy with the OZI rule
in these reactions was found by a factor of 10-100.

        How is it possible to coincide the nice agreement of the OZI rule
predictions with results of some experiments and its
complete failure in the others?
The main idea of the explanation is that the OZI rule itself is
always valid. Some deviations from this rule are only apparent. They are
due to non-trivial dynamics of the process, which can not be
described by a diagram with disconnected quark lines.

For instance, the OZI rule forbids the production
of a pure \s~state in the nucleon-nucleon interaction,
if there are no strange quarks in the nucleon.
Only in this case the production of the \s~ pair is described by a
disconnected diagram.
But if the strange quarks in the nucleon play a non-negligible role, then
the \s~ pair could be produced in the nucleon-nucleon interaction
via shake-out or rearrangement of the strange quarks already stored
in the nucleon. This process is described by a connected
diagram and the OZI rule suppression is not applicable in this situation.
Therefore the observed violation of the OZI rule is only apparent.
It reflects that the \f~mesons may be created from the strange quarks
already stored in the nucleon.

        However this line of arguments immediately induces new
questions. Why does the degree of the OZI rule violation depend
drastically on the quantum numbers of the initial state?
Why is the OZI rule violated in the \an~ of \ap~ at rest, but not
in flight?
Why  the strong violation of the OZI rule
has been observed till now only in some channels  but not in all  reactions of
 $\pi p$, $NN$ and $\bar{N}N$ interactions?

        To answer these questions, a model of polarized
nucleon strangeness
was proposed
\cite{Ell.95}, \cite{Ell.99}.
The main idea was suggested by the results of the experiments on
lepton deep inelastic scattering on polarized targets
\cite{EMC.88}, \cite{SMC.97}, \cite{E143},\cite{E155}
which may be considered as an indication
on the polarization of the strange quarks in the nucleon sea. It turns
out that the polarization of the nucleon strange sea may naturally
explain the observed dependence of the degree of the OZI rule
violation from the initial state quantum numbers, initial energy and
the final state content.
A number of tests of the model was proposed
\cite{Alb.95}, \cite{Ell.95b},
part of them was already successfully confirmed by the experiment.

        In this review we consider first the
phenomenology of the OZI rule (Section 2). The experimental
evidences of the large OZI violation are discussed in Section 3.
Section 4 is devoted to the polarized strangeness model and its
explanation of the large OZI violation. In Section 5 a review of
other theoretical models is given. Section 6 contains a summary
and discussion of the future experiments.

\section{The OZI rule}

\subsection{Phenomenology of the OZI rule}

There are different formulations of the OZI rule (for review, see
\cite{Lip.91}). On the phenomenological level it is worthwhile to
consider the relevant physics following the approach
of Okubo \cite{Oku.77}.

       Let us consider creation of
$q\bar{q}$ states in the interaction of hadrons
\begin{eqnarray}
A+B\longrightarrow C + q\bar{q}, ~~~~\mbox{for q=u,d,s}
\end{eqnarray}
where hadrons $A,B$ and $C$ consist of only light quarks.

The OZI rule in the formulation of Okubo demands that
\begin{eqnarray}
Z = \frac{\sqrt{2} M(A+B\rightarrow C+s\bar s)}
{M(A+B\rightarrow C+u\bar u) + M(A+B\rightarrow C+d\bar d)}=0 \label{OZI}
\end{eqnarray}
where $M(A+B\rightarrow C+q\bar{q})$ are the amplitudes of the
corresponding processes.

It means that if the $\phi$ meson is a pure $\bar{s}s$ state,
it could not have been produced in the interaction of ordinary hadrons.
The OZI rule in Okubo's form strictly forbids creation of
strangeonia or charmonia in the interaction of hadrons composed
from $u$ and $d$ quarks only.

\begin{figure}[ht]
\setlength {\unitlength} {1mm} \thicklines
\begin{picture}(90,70)(0,10)
\put(10,30){\line(1,0){20}}

\put(10,35){\line(18,0){18}}

\put(10,40){\line(15,0){15}}


\put(10,60){\line(1,0){35}}

\put(10,55){\line(18,0){18}}

\put(10,50){\line(15,0){15}}


\put(28,55){\line(0,-1){20}}

\put(25,50){\line(0,-1){10}}

\put(30,30){\line(0,+1){25}}


\put(30,55){\line(1,+0){15}}

\put(35,30){\line(1,+0){10}}

\put(35,35){\line(1,+0){10}}

\put(35,30){\line(0,+1){5}}

\put(47,34){$\bar{s}$}
\put(47,29){$s$}
\put(52,32){$\phi$}
\put(52,60){$\pi$}
\put(1,30){$N$}
\put(1,60){$\bar{N}$}
\put(7,30){$u$}
\put(7,35){$u$}
\put(7,40){$d$}
\put(7,60){$\bar{u}$}
\put(7,55){$\bar{u}$}
\put(7,50){$\bar{d}$}
\put(47,60){$\bar{u}$}
\put(47,55){$u$}



\put(70,30){\line(1,0){20}}

\put(70,35){\line(1,0){18}}

\put(70,40){\line(1,0){15}}


\put(70,60){\line(1,0){35}}

\put(70,55){\line(1,0){18}}

\put(70,50){\line(1,0){15}}


\put(85,50){\line(0,-1){10}}


\put(90,55){\line(1,+0){15}}

\put(90,30){\line(0,+1){25}}


\put(88,35){\line(1,0){17}}

\put(88,55){\line(0,-1){15}}

\put(88,40){\line(1,0){17}}


\put(110,35){$\omega$}

\put(110,60){$\pi$}



\put(65,30){$u$}
\put(65,35){$u$}
\put(65,40){$d$}
\put(65,60){$\bar{u}$}
\put(65,55){$\bar{u}$}
\put(65,50){$\bar{d}$}
\put(105,60){$\bar{u}$}
\put(105,55){$u$}
\put(105,40){$\bar{u}$}
\put(105,35){$u$}
\put(25,20){ a)}
\put(85,20){ b)}
\end{picture}
\caption{Quark diagrams for the
$\phi$ (a) and
$\omega$ (b) - meson production in
$\bar{N}N$ \an.}
\label{diag}
\end{figure}
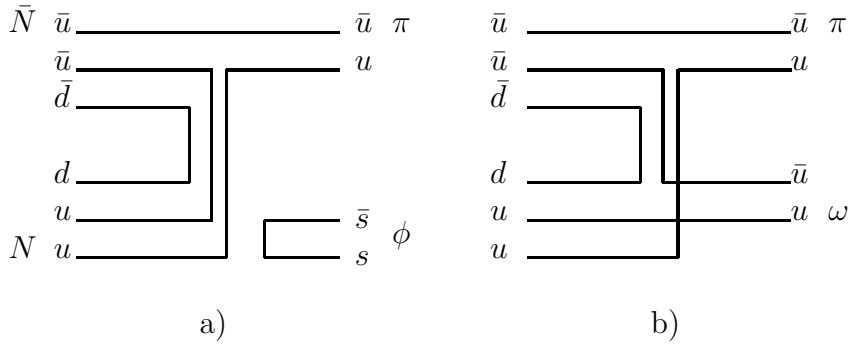

Sometimes the OZI rule is stated as a suppression of
reactions with disconnected quark lines.
Let us consider creation of the \f~meson,
assuming that it is a pure \s~state.
If there are no strange quarks in the nucleon, then the production
of the \f~meson, for instance, in the \p~\an,
should be described by the diagram in Fig.
\ref{diag}a.
The quark lines of the
\s~ pair in the final state are not connected with the
quark lines of the initial state and according to the OZI
rule this reaction should be suppressed in comparison with the
production of the \om~meson. The diagram of the \om~meson production
is shown in Fig. \ref{diag}b.

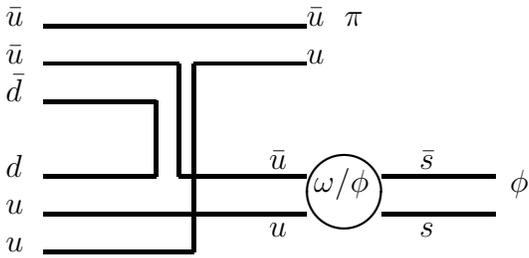
\begin{figure}[ht]
\setlength {\unitlength} {1mm} \thicklines
\linethickness{0.5mm}
\begin{picture}(90,70)(0,10)


\put(10,30){\line(1,0){20}}

\put(10,35){\line(1,0){18}}

\put(10,40){\line(1,0){15}}


\put(10,60){\line(1,0){35}}

\put(10,55){\line(1,0){18}}

\put(10,50){\line(1,0){15}}


\put(25,50){\line(0,-1){10}}


\put(30,55){\line(1,+0){15}}

\put(30,30){\line(0,+1){25}}


\put(28,35){\line(1,0){17}}

\put(28,55){\line(0,-1){15}}

\put(28,40){\line(1,0){17}}

\put(55,35){\line(1,0){15}}
\put(55,40){\line(1,0){15}}
\put(46,38){$\omega/\phi$}
\put(50,38){\circle{10}}
\put(50,60){$\pi$}



\put(5,30){$u$}
\put(5,35){$u$}
\put(5,40){$d$}
\put(5,60){$\bar{u}$}
\put(5,55){$\bar{u}$}
\put(5,50){$\bar{d}$}
\put(45,60){$\bar{u}$}
\put(45,55){$u$}
\put(40,41){$\bar{u}$}
\put(40,32){$u$}
\put(60,41){$\bar{s}$}
\put(60,32){$s$}
\put(72,38){$\phi$}


\end{picture}
\caption{Quark diagram of the OZI allowed mechanism for the
$\phi$ production in
$\bar{N}N$ \an.}
\label{allow}
\end{figure}

The only way to form \f,~ according to the OZI rule, is through admixture
of the light quarks in the \f~ wave function. The corresponding diagram is
shown in Fig. \ref{allow}.
The admixture of light quarks appears in the \f~ wave function because
the $\phi$ and $\omega$ mesons are mixture of
SU(3) singlet $\omega_0$ and octet $\omega_8$ states:
\begin{eqnarray}
\phi = \cos{\Theta}~\omega_8 - \sin{\Theta}~\omega_0\\
\omega = \sin{\Theta}~\omega_8 + \cos{\Theta}~\omega_0
\end{eqnarray}
where
\begin{eqnarray}
\omega_8 = (u\bar{u} + d\bar{d} - 2 s\bar{s})/\sqrt{6} \label{8}\\
\omega_0 = (u\bar{u} + d\bar{d} +  s\bar{s})/\sqrt{3} \label{0}
\end{eqnarray}

Therefore the \f~ wave function is

\begin{equation}
\phi =
(u\bar{u} + d\bar{d}) (-\sin{\Theta}\frac{1}{\sqrt{3}}+
\cos{\Theta} \frac{1}{\sqrt{6}} )-
 s\bar{s} (\sin{\Theta}\frac{1}{\sqrt{3}} + \cos{\Theta}\frac{2}{\sqrt{6}})
\label{mix}
\end{equation}

Introducing the ideal
mixing angle  $\Theta_i=35.3^0$, for which
\begin{equation}
\cos{\Theta_i}=\sqrt{\frac{2}{3}},~~~\sin{\Theta_i}=\sqrt{\frac{1}{3}}
\end{equation}
one may rewrite (\ref{mix}) as follows:
\begin{eqnarray}
\phi &= - \cos{(\Theta_i - \Theta)}~|\bar{s}s> +
\sin{(\Theta_i - \Theta)}~|\bar{q}q> \\
\omega &= \cos{(\Theta_i - \Theta)}~|\bar{q}q> +
\sin{(\Theta_i - \Theta)}~|\bar{s}s>
\end{eqnarray}
where $|\bar{q}q> = \frac{1}{\sqrt{2}}(\bar{u}u + \bar{d}d)$.\\

If  $\Theta = \Theta_i $,
then \f~ is a pure \s~ state.

 However the physical mixing
angle $\Theta$ differs from the ideal one. It is determined by the masses of
the mesons in the corresponding
nonet. For the vector mesons the physical mixing angle is

\begin{equation}
\tan^2{\Theta} = \frac{m^2_{\phi} - m^2_{\omega_8}}{m^2_{\omega_8} -
m^2_{\omega}} \label{z1}
\end{equation}
where  from the quadratic Gell-Mann-Okubo
mass formula

\begin{equation}
m^2_{\omega_8}= \frac{4 m^2_{K^*} - m^2_{\rho}}{3} \label{z2}
\end{equation}

        Substituting the masses of the vector mesons in
eqs.(\ref{z1})-(\ref{z2}), one obtains $\Theta=39^0$, which is not far from
the ideal mixing angle.
The difference $\Theta -\Theta_i$ determines the
contribution of light quarks in the
$\phi$ wave function. This contribution,  according to the OZI--rule,
determines how large the cross sections of \f~
production are in $NN$, $\pi N$ or $\bar{N}N$
interactions. To demonstrate this, let us rewrite
eq.~(\ref{OZI}) in terms of
mixing angles:

\begin{eqnarray}
\frac{M(A+B\rightarrow C+\phi)}{ M(A+B\rightarrow C+\omega)} =
- \frac{ Z+ \tan(\Theta-\Theta_i)}{1-Z\tan(\Theta-\Theta_i)} \label{MAB}
\end{eqnarray}

 If the OZI rule is correct, i.e. the parameter $ Z$ is $Z=0$,
then
\begin{equation}
\frac{M(A+B\rightarrow C+\phi)}{ M(A+B\rightarrow C+\omega)} =
 - \tan(\Theta-\Theta_i) \label{mr}
\end{equation}

and
\begin{equation}
R = \frac{\sigma(A+B \rightarrow\phi X)}{\sigma(A+B\rightarrow\omega X)}
=\tan^2{(\Theta - \Theta_i)} \cdot f \label{R}
\end{equation}

where $f$ is the ratio of phase spaces of the reactions.
If $f=1$ and $\Theta=39^0$, then the OZI rule predicts
that in all hadron reactions the ratio between the cross sections of \f~ and
\om~ production $R(\phi/\omega)$
should be:

\begin{equation}
R(\phi/\omega) = 4.2\cdot10^{-3}
\label{rfi}
\end{equation}

Therefore, using as input only the masses of mesons, the ratio of
production cross sections is predicted.

It is remarkable to what
extent the experimental data in different interactions and at
different energies
follow this rule. We will discuss this in detail in the next sections.

The physical mixing angle could be calculated for each meson nonet using the
quadratic Gell-Mann-Okubo mass formula of eqs.(\ref{z1})-(\ref{z2}).
It gives for the tensor mesons  $\Theta=28^0$ and for the $3^{--}$ nonet
- $\Theta=29^0$ \cite{PDG}. In both cases the deviation from the ideal
mixing angle is not large. It means that the mixing is not large and
the production of the corresponding \s~states will be suppressed in comparison
with their light quark counterparts.
The small mixing is predicted \cite{Gei.91} also for the
nonets with $J^{PC}=1^{++},1^{+-}$
and $4^{++}$.

Let us consider, for instance, the tensor nonet.
The tensor meson \s~ state is $f_2'(1525)$ meson and
its partner made of light quarks is $f_2(1270)$.
The OZI rule
predicts that the ratio $R(f_2'(1525)/f_2(1270))$ is:

\begin{equation}
R=\frac{\sigma(A+B \rightarrow f_2'(1525) + X)}
{\sigma(A+B\rightarrow f_2(1270) + X)}  = 16\cdot10^{-3}
\label{rf2}
\end{equation}
The space factor is assumed to be f=1.

 The situation with the mixing of the pseudoscalars
is special (see, \cite{Leu.98}, \cite{Fel.99},\cite{Fel.02}).
It reflects a special
role of the pseudoscalars as the Goldstone bosons connected with
spontaneous breaking of the chiral symmetry. The non-perturbative
QCD effects are essential for the mass spectrum of the pseudoscalars,
especially for the mass of $\eta'$ meson. In result, the mass spectrum of
pseudoscalars differs from the mass
spectrum of
"normal" mesons.

Phenomenologically this deviation manifests itself as
large deviation from the ideal mixing angle which is
$\Theta - \Theta_i \approx -(45-55)^0$.
The values of the mixing angle from $\Theta=-10^0$ to $\Theta=-23^0$ have been
obtained in different analysis. Later we will discuss the situation
with the mixing of pseudoscalars in more
details. The only remark here is the following.

Let us take recent result by the KLOE collaboration \cite{Klo.01}
that
the pseudoscalar mixing angle is
$\Theta= (-14.7^{+1.7}_{-1.5})^0$. Then
the ratio between
the cross sections of $\eta$ and  $\eta'$ production is
\begin{equation}
R= \frac{\sigma(A+B \rightarrow\eta X)}{\sigma(A+B\rightarrow\eta' X)}
\sim 1.42 \label{reta}
\end{equation}

        Therefore, large mixing induces $\eta$ dominance over $\eta'$
in contrast with the \f~suppression over \om~ induced by the
small mixing of the vector mesons.

Strong mixing is expected \cite{Isg.00},\cite{Ani.97}
for the scalar mesons with $J^{PC}=0^{++}$.
However, the very content of the scalar nonet is
still under discussion now (see,e.g. \cite{PDG}).

Let us consider, how the existing experimental data correspond to
the predictions of
eqs. (\ref{rfi}), (\ref{rf2}) and (\ref{reta}).

\subsection{The OZI rule for the vector mesons production}

        The production of the
\f~ and \om~ mesons was studied in different experiments in $NN$,
$\pi N$ and $\bar{N}N$ interaction. The obtained ratios $R(\phi X/\omega X)$
of the cross sections of \f~ and \om~production
 and the values of the OZI--rule violation
parameter $Z$ are shown
in the Table \ref{tpp}. The parameter $Z$ was calculated for
$\delta=\Theta-\Theta_i=3.7^0$, assuming the same phases of the
\f~ and \om~ production amplitudes.

From these results one may obtain that in $\pi N$ interactions
the weighted average ratio of the cross sections of \f~ and \om~production
at different energies is

\begin{equation}
\bar{R}(\frac{\sigma(\pi N\to \phi X)}{\sigma(\pi N\to \omega X)})
= (3.30\pm0.34)\cdot10^{-3}
\label{eq:RV1}
\end{equation}

It corresponds to the value of the OZI violation parameter
$ |Z_{\pi N}| = 0.6\pm0.3 \%$. Therefore, in $\pi N$ interaction the
agreement with the OZI rule prediction (\ref{rfi}) is practically perfect.

 The weighted average ratio of the cross sections of the \f~ and
 \om~production
at different energies
in nucleon-nucleon interactions is not far from the OZI "magic"
value of (\ref{rfi}) either:

\begin{equation}
\bar{R}(\frac{\sigma(NN \to \phi X)}{\sigma(NN\to \omega X)})
= (12.78\pm0.34)\cdot10^{-3}
\label{eq:RV2}
\end{equation}
It corresponds to the value of the OZI violation parameter
$ Z_{NN} = 6.0\pm0.2 \%$.

        Similar values are found for the \ap~\an~ in flight:
\begin{eqnarray}
\bar{R}(\frac{\sigma(\bar{p}p \to \phi X)}{\sigma(\bar{p}p\to \omega X)})
 &= & (14.55\pm1.92)\cdot10^{-3}\\
 Z_{\bar{p}p} & = & 5.8\pm0.8 \% \nonumber
\end{eqnarray}

        The comparison of the above results for $\pi N$ scattering
with those
for $NN$ and  $\bar{N}N$ put in evidence that:
\begin{itemize}
\item The deviation from the OZI rule prediction in the  $NN$ and  $\bar{N}N$
scattering is significantly higher than in the $\pi N$ interaction.
\item In general the deviation from the OZI rule for
the vector meson production does not exceed 10$\%$.
\end{itemize}

        One should note that no correction for the phase space
difference of the reaction final state was applied.
        A correct procedure to test the OZI rule needs the comparison
of not the cross sections but
the amplitudes of the \f~ and \om~ productions. It was done in \cite{Cas.xx}
for $\pi N$ and $NN$ scattering. Their conclusions coincide in general
with ours:
the OZI rule for the vector mesons production  is valid within 10$\%$
accuracy.

\subsection{The OZI rule for the tensor meson production}

 The experimental data on the tensor meson production are more
scarce, but in general they are also confirm the OZI rule prediction
(\ref{rf2}).

        There are some specific features in the tensor meson production
which should be kept in mind. First, the determination  of the
tensor meson mixing angle directly from the masses of the mesons is not
a persuading procedure due to the large width of the tensor mesons.
Therefore, it is worth checking the numerical prediction of
eq.(\ref{rf2}) in another way, by
using the data on the
decay widths of the \ten~ into  $\pi\pi$ and $\bar{K}K$
channels. Taking these data from \cite{PDG}, one may obtain
the ratio of the  \ten~ decay widths into
the OZI-forbidden
$\pi\pi$ and OZI-allowed $\bar{K}K$ modes, corrected on the phase space
difference:

\begin{equation}
R=\frac{W(f_2'(1525)\to \pi\pi)}{W(f_2'(1525) \to \bar{K}K)} =
(2.6\pm0.5)\cdot10^{-3}
\label{eq:RTdecay}
\end{equation}

This value is slightly less than the eq.(\ref{rf2}) one.
It indicates that the
suppression of the \ten~ on the $f_2(1270)$ should be at the level of
\begin{equation}
R=\frac{\sigma(A+B \rightarrow f_2'(1525) + X)}
{\sigma(A+B\rightarrow f_2(1270) + X)}  =
(3-16)\cdot10^{-3} \label{tns}
\end{equation}

        Another complication of the analysis of the tensor meson production
is related to the fact that it is quite hard
to select correctly the contribution of the $f_2(1270)$ meson from
nearby the $a_2(1320)$ meson.
Frequently the total contribution of  the $f_2(1270)$
and $a_2(1320)$ mesons is given.

 In Table \ref{tf2} the ratio of the cross sections
$R=f_2'(1525) X/f_2(1270) X $ is given for different reactions.

        One can see that the general agreement with  the OZI rule prediction (\ref{tns})
        is quite well. There are some experiments with large  $R=f_2'(1525) X/f_2(1270) X $
        ratios, but  the errors are also
large to be conclusive. Averaging the results for the \ap~\an~ gives

\begin{eqnarray}
\bar{R}(\frac{\sigma(\bar{p}p \to f_2'(1525) X)}{\sigma(\bar{p}p\to f_2(1270) X)})
 &= & (27\pm6)\cdot10^{-3}\\
 Z_{\bar{p}p} & = & 3.9\pm1.6 \% \nonumber
\end{eqnarray}

This makes  even more interesting the result obtained by the OBELIX
collaboration \cite{Pra.98}. It was demonstrated that the large violation of the
OZI rule occurs in some specific condition: when
annihilation takes place from the P-wave. We will discuss it in detail
in Sect.3.1.4.

\subsection{The OZI rule for the pseudoscalar mesons production}

The mass spectrum of pseudoscalar mesons is
not similar to the spectrum of ordinary mesons.
The non-perturbative
QCD effects are of most importance just in this sector. They lead
to the strong upward shift of the $\eta'$ meson mass.
For normal mesons the pure SU(3) octet state $\omega_8$
from (\ref{8}) is always heavier than the singlet $\omega_0$ state
from (\ref{0}). It is quite natural because  by definition
(\ref{8})-(\ref{0})
the contribution of strange quarks is larger for  $\omega_8$.
However for the pseudoscalars, the state $\eta_8$ with more strangeness
content is
lighter than the corresponding  $\eta_0$ partner.
So, the mixing between \s~  and light $\bar{q}q$ states is large
and in this sense
the OZI rule for the pseudoscalar sector is strongly
violated.

The complications related to  a special status of the
pseudoscalars is not ended by a large value of the mixing angle.
H.Leutwyler \cite{Leu.98} has shown that two mixing angles
$\theta_8$ and $\theta_0$ are needed for accounting different $SU(3)_f$
violation for the octet and singlet flavor states. The values of the mixing
angles are
$\theta_8 \sim -20^0,~\theta_0 \sim -4^0$. Correspondingly, two different
coupling constants $f_8$ and $f_0$ are needed .

Feldmann, Kroll and Stech \cite{Fel.99} have introduced a quark flavor
basis instead of the octet-singlet one. In this approach $\eta$ and
$\eta'$ are described as linear combinations of orthogonal states
$\eta_q$ and $\eta_s$ with the flavor structure
$q\bar{q}=(u \bar{u} + d \bar{d})/\sqrt{2}$ and $ s \bar{s}$, respectively.

In this scheme it is possibile to reduce
four constants to the single mixing angle $\Phi$ and two coupling
constants $f_q$ and
$f_s$.
The analysis of a number of decay and scattering processes
allows to fix these parameters. The ratio of cross sections
for $\eta$ and $\eta'$ production at high energies where
phase space corrections assumed to be non-significant is
\begin{equation}
R=\sigma(A+B \to \eta +X)/\sigma(A+B \to \eta'+X) = \tan^{-2}\Phi
\end{equation}

According to \cite{Fel.99} the value of $\Phi$ is $\Phi=(39.3\pm1)^0$
It leads to
the following ratio between
the cross sections of the $\eta$ and $\eta'$-meson production

\begin{equation}
R (\eta/\eta') \sim 1.49 \label{reta1}
\end{equation}

It should be compared with the standard, one-mixing-angle scheme
prediction of eq.(\ref{reta}) that $R (\eta/\eta') \sim 1.42$.

The ratios between the cross section of the $\eta$ and  $\eta'$-mesons
in different reactions are collected in Table \ref{teta}.

        In spite of the delicate situation with the mixing of the
pseudoscalars,
a superficial look at Table
\ref{teta} provides an
impression that the bulk of the data are in agreement with the
predictions of eqs.(\ref{reta}) and (\ref{reta1}).

        However,
there are some marked deviations from the OZI rule prediction of
eq.(\ref{reta}) found in the measurements of $\eta$ and $\eta'$ production
in the proton-proton interaction in the vicinity of their thresholds
\cite{Cal.96}, \cite{Mos.99}. (For review of the experimental situation,
see \cite{Mos.00}). It turns out that, for instance, at the proton energy
Q=2.9 MeV above the threshold the ratio of cross sections is
$R (\eta/\eta') = 37.0\pm11.3$. At
Q=4.1 MeV  the ratio is
$R (\eta/\eta') = 26.2\pm5.4$.

A standard explanation (see, e.g. \cite{Ger.90})
of the reason of this anomaly is
increasing of the $\eta$ production due to the nucleon $S_{11}(1535)$ resonance
 which
strongly couples with  the $N \eta$ final state. The excitation of this
resonance is essential just near the $\eta$ production threshold.
However, the quantitative results depend on the
$S_{11}(1535)$
 coupling constants
which are not well known. Moreover, to provide large ratio
$R (\eta/\eta')$  similar resonances should be absent
near the $\eta'$ production threshold. However, that may be not the case,
as recent calculations \cite{Nak.99} have shown
that the excitation of analogous  $N^*$ resonances, such as $S_{11}(1897)$ and
$P_{11}(1986)$,  could reproduce the $\eta'$ cross threshold near
the threshold.

\subsection{Why does the OZI rule work?}

        Quite often the OZI rule is formulated as suppression of the
processes described by the disconnected diagrams.
        Natural explanation of this hierarchy is provided in the
large  $N_c$  limit of QCD \cite{Wit.79}, \cite{Ros.80}.
It has been shown that at large $N_c$  the diagrams with the smallest
possible number of quarks loops dominate because the QCD coupling
constant tends to zero if $N_c$ is sent to infinity. Therefore
the disconnected
diagrams are suppressed by higher powers of $1/N_c$ compared with the
connected ones.

        However in the real world with $N_c=3$ the degree of the
suppression of the OZI violating processes is much stronger than 1/3.
As we discussed in the Sect. 2.2 and 2.3 for the vector and tensor
meson production cross sections this suppression factor is of the order
of $10^{-3}$.

        On a phenomenological level the reasons of applicability
of the OZI rule was thoroughly analysed by H.Lipkin and also
by P.Geiger and N.Isgur
(see \cite{Lip.91},\cite{Gei.91},\cite{Isg.00},\cite{Gei.97}
and references therein).
They pointed out the importance
of  cancellations between different intermediate states as a reason for
the validity of the OZI rule.

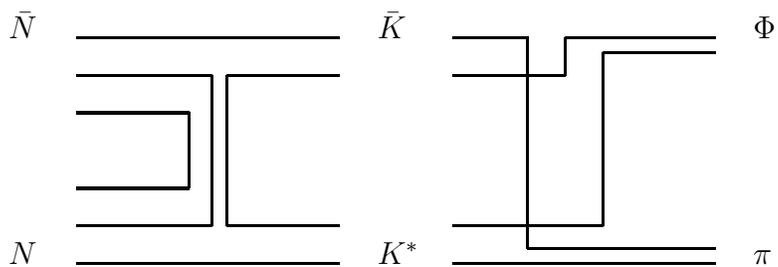
\begin{figure}[ht]
\setlength {\unitlength} {1mm} \thicklines
\begin{picture}(90,70)(0,10)
\put(10,30){\line(1,0){35}}
\put(10,35){\line(18,0){18}}
\put(10,40){\line(15,0){15}}
\put(10,60){\line(1,0){35}}
\put(10,55){\line(18,0){18}}
\put(10,50){\line(15,0){15}}
\put(28,55){\line(0,-1){20}}
\put(25,50){\line(0,-1){10}}
\put(30,35){\line(15,0){15}}
\put(30,55){\line(15,0){15}}
\put(30,55){\line(0,-1){20}}
\put(60,60){\line(1,0){10}}
\put(60,55){\line(1,0){15}}
\put(60,35){\line(1,0){20}}
\put(60,30){\line(1,0){35}}
\put(70,60){\line(0,-1){28}}
\put(70,32){\line(1,0){25}}
\put(75,55){\line(0,1){5}}
\put(75,60){\line(1,0){20}}
\put(80,35){\line(0,1){23}}
\put(80,58){\line(1,0){15}}

\put(100,60){$\Phi$}
\put(100,30){$\pi$}
\put(1,30){$N$}
\put(1,60){$\bar{N}$}
\put(50,30){$K^*$}
\put(50,60){$\bar{K}$}

\end{picture}
\caption{Annihilation  $\bar{p}p \to \phi \pi$
as a two-step process.}
\label{tstep}
\end{figure}

Indeed, all OZI--violated
reactions could be regarded as two-step processes, for instance:
\begin{eqnarray}
\phi &\to \bar{K} K \to \rho \pi  \label{phi1}\\
\bar{p} p & \to \bar{K} K^* \to \phi \pi \\
\pi^- p & \to  K^0 \Lambda \to \phi  n   \label{phi2}
\end{eqnarray}
At each
step the process is described by a connected OZI-allowed diagram (see, Fig. \ref{tstep}) .
There is
no suppression for each sub-process. Why
then  the total process is suppressed?

H.Lipkin \cite{Lip.87} argued that in some sense this suppression
could be considered as a reflection of the underlying flavor symmetry
which
"equalizes" the contributions of different intermediate states.
To demonstrate the cancellation between these contributions,
he introduced an analog of the G-parity in the case
of the total flavor symmetry.
Under conventional G-parity the
$u$-quark is transformed into $\bar{d}$ (as well as $d$ into $\bar{u}$).
The corresponding transformation in the case of flavor symmetry
transforms light and strange quarks. It
interchanges
$u$ and $\bar{s}$ (as well as $s$ and $\bar{u}$). The strong interaction and
the T-matrix are invariant under this general G-parity in the case of
flavor symmetry. Then one could classify the meson states into
odd or even eigenvalues of the generalized G-parity.
It was shown in \cite{Lip.87} that the contributions from intermediate
states having even and odd eigenvalues of the generalized G-parity have
opposite phases. So the physical reason for the validity of the OZI rule
is the cancellation of the contributions from different intermediate states
in the transition amplitude of two-step processes.

        If this delicate cancellation does not occur for some
reasons,  e.g. near the thresholds where some intermediate states are
open but others are closed, one may expect to observe the deviation
from the OZI rule predictions. Also if a theoretical analysis considers
only a few diagrams, it may easily results in a substantial OZI rule
violation.

This conclusion is very important because a lot of models claimed to
reproduce the large violation of the OZI rule, but considering only a very limited set of diagrams
of few meson exchanges. So it is not a surprise that the resulting
non-compensation allows to obtain a large OZI rule violation.

        Concrete calculations made by
Geiger and Isgur \cite{Gei.97} have shown strong cancellations
between mesonic loop diagrams consisting of S-wave strange mesons
($\bar{K}K, \bar{K}K^*,  \bar{K}^*K^*$) with loops containing
$K$ or $K^*$ plus one of the four strange mesons of the $L=1$ nonet.
To maintain the cancellation, a summation of very large set,
up to ten thousand (!) channels,  is needed \cite{Isg.00}.

        The development of the QCD provides us more understanding of the
applicability of the OZI rule. One of the first success of the
QCD sum rules approach \cite{Shi.79} was the correct calculation of
the value of \f~-\om~
mixing. It was shown \cite{Shi.79} that the reason
of the smallness of the mixing
is the suppression of contribution of the vacuum
intermediate state in the matrix elements
$<0|\bar{q}\Gamma q \bar{q}\Gamma q|0>$ , where $\Gamma$ are some
matricies acting on colour, flavor and spinor indices.

\begin{figure}[ht]
\begin{center}
\epsfig{file=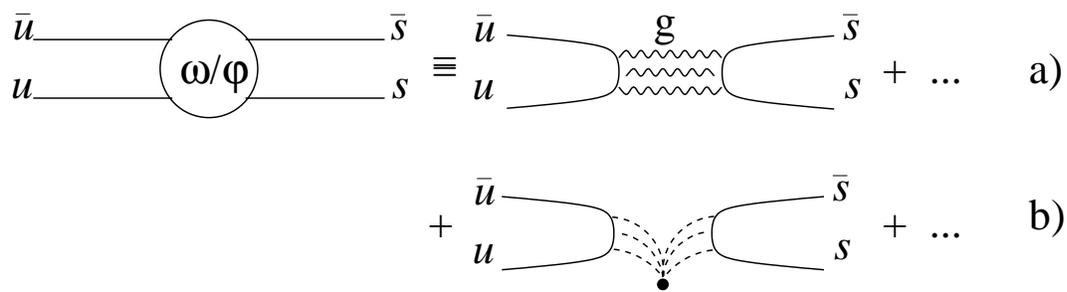,width=14.cm}
\caption{Quark diagrams of the \f-\om~mixing.
a) gluon exchange, b) mixing in the instanton
field. The instanton is shown by the black point.}
\label{trans}
\end{center}
\end{figure}

        Indeed, let us consider in more details what are the
consitutents of the blob in Fig. \ref{allow} which describes the
$\phi$ production in \p~ \an.
According to the OZI rule we should consider this process as a
two step process: the first step is the formation of the light quark
$\bar{q}q$ pair with effective mass of \f-meson, the second step is the
transition of the $\bar{q}q$ pair to the \s~pair. This transition
schematically depicted by the diagrams of Fig. \ref{trans}
could be described either via gluon exchange (see, Fig.\ref{trans} a)
or via some non-perturbative processes (Fig.\ref{trans} b).

Studies of the three-gluon exchange diagram of Fig.\ref{trans} a)
in the $\phi \to \rho \pi$
decay  have shown \cite{Ara.77},\cite{Ges.80} that its contribution is
small and of wrong sign. Non-perturbative effects of Fig.\ref{trans} b)
are dominating. This assumption leads to very non-trivial dependence
of the OZI rule violation on the quantum numbers of investigated channel.
It was shown in \cite{Ges.80} that in the dilute instanton gas
approximation the non-perturbative processes of Fig.\ref{trans} b) are suppressed
for the vector and tensor channels but non-vanishing for the axial vector,
pseudoscalar and scalar channels. Qualitatively it fits with the
observed picture that just in the pseudoscalar
and scalar sectors
 the meson mass spectra are not
similar to the conventional one, obtained with small deviation from the ideal
mixing.

\begin{figure}[ht]
\begin{center}
\epsfig{file=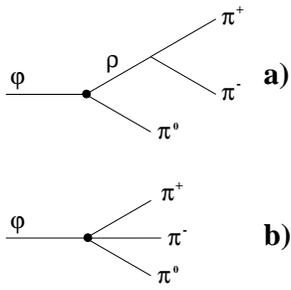,width=8.cm}
\caption{The diagrams of the $\phi \to \pi^+ \pi^- \pi^0$ decay.
a) via
 \f-$\rho$~mixing b) direct transition to $3\pi$ system.}
\label{mixing}
\end{center}
\end{figure}

        Of course, some direct OZI  violation is possible.
The direct OZI violation means the violation in the amplitude of the
\f~ production or decay. Indeed, till now we
assume that the
\f~mesons are produced  (and decayed) {\it only} via
$\phi - \omega$ or $\phi - \rho$ mixing, i.e. due to
light quark components in the \f~ wave function. An example
of the mixing is shown in Fig. \ref{mixing} a) for the case of
the \f~ decay into $\pi^+ \pi^- \pi^0$. However
the diagram of Fig. \ref{mixing} b), which describes direct
transition of \f~ into three-pions system, is also possible.
Similar situation exists in the neutral kaon systems, where
there is a distinction between the direct CP-violation and
CP-violation due to the mixing of $K_S$ and $K_L$ mesons.

        The role of mixing and direct transitions in the \f~ decays
was analysed in \cite{Ach.92}, \cite{Ach.96}, \cite{Ach.00}.
Recent experiments at VEPP-2M
collider have provided some indications  of the direct
OZI violation \cite{Ach.99,AchM.00}. The KLOE collaboration
at DA$\Phi$NE \f-factory has claimed
on the finding direct OZI violation in $\phi \to \pi^+ \pi^- \pi^0$ decay
\cite{Klo.00}. The amplitude of direct transition turns out to be
as large as
10\% of the $\rho \pi$ one.

Summarizing it is possible to say that the OZI rule reflects
an important feature of the hadron interactions -
suppression of the flavor mixing transitions.
It reflects the absence of the
processes with pure gluonic intermediate
states. One should stress that when the gluonic
intermediate states are important, the OZI rule in the Okubo's
form of eq.(\ref{OZI}) is not valid. In that case the \s~ pair production is not
suppressed as large as the mixing angle prescriptions
of eqs.(\ref{rfi}),(\ref{rf2}),(\ref{reta}). A nice example is
the  decays of
$J/\psi \to \phi + X$ and $J/\psi \to \omega + X$. The both decays are
OZI-forbidden, however, there is no substantial \f~meson suppression as
predicted by the mixing angle formula (\ref{rfi}). Indeed, the
ratios of \f~ and \om~final states \cite{PDG} are rather similar and far from the
$R(\phi/\omega)=4.2\cdot10^{-3}$ prediction of eq.(\ref{rfi}):
\begin{eqnarray}
R_{\pi\pi}=\frac{W(J/\psi \to \phi +\pi^+ \pi^-)}{W(J/\psi \to \omega+\pi^+ \pi^-)} =
(111\pm23)\cdot 10^{-3} \\
R_{\eta}=\frac{W(J/\psi \to \phi +\eta)}{W(J/\psi \to \omega+\eta)} =
(411\pm61)\cdot 10^{-3} \\
R_{f_0(980)}=\frac{W(J/\psi \to \phi +f_0(980))}{W(J/\psi \to \omega+f_0(980))} =
(2290\pm1040)\cdot 10^{-3}
\end{eqnarray}

The value of the flavor mixing is channel-dependent. It
is large for the
pseudoscalar and scalar channels. For other channels the OZI
rule is a nice approximation.
As it was discussed in \cite{Ven.90}, the OZI limit
of QCD is a more accurate approximation than the large $N_c$ limit,
or the quenched approximation, or the topological expansion
($N_c \to \infty$ at fixed
$N_f/N_c$).

\section{Apparent violation of the OZI--rule}

        The previous Section demonstrates that the OZI--rule
predictions have been tested many times in different reactions and
 the general conclusion is
that the rule is
working quite well and is valid within 10$\%$
accuracy. It is a remarkable agreement, bearing in mind that it is
based only
on the value of the mixing angle, i.e. on the values of meson masses.
It is also impressive that the agreement
is valid at different energies from 100 MeV till 100 GeV.

        In this situation it was a surprise when,
in spite of the solid theoretical background and numerous experimental
confirmations, the experiments at
LEAR (CERN) with stopped \ap s showed large violations of the OZI
rule
(for a review, see  \cite{Ams.98}, \cite{Sap.98}).
In some \an~ reactions the deviation from the OZI--predictions was
as large as a factor of 30-70.

Then it was established that this anomaly was not restricted only to the
antiproton \an~at rest but there were similar deviations
in $\pi N$, $pp$ and $pd$ interactions. How should we treat these
experimental evidences?

        We would like to advocate the point of view that the OZI
rule itself is always valid. In cases where some deviation from the
OZI rule predictions exists, it should be regarded as a signal on
non-trivial physics, as a signal that the dynamics of the processes is
more complicated than expected. For instance, let us assume,
following \cite{Ell.95}, that
the nucleon wave function contains \s~pairs which might take part
in the  \f~ production in $NN$ or $\bar{N}N$ interactions.
Then such processes would be described by connected diagrams and
therefore would not be  OZI-suppressed.
So the violation of the OZI rule observed in these processes is only
apparent.

        A similar ideology was used to search for exotic mesons
and baryons \cite{Lan.99}. An exotic 4--quark meson $\bar{q}q\bar{s}s$
should decay more readily into $\phi X$ system demonstrating an apparent
violation of the OZI rule. Therefore, the apparent violation of the OZI rule
give us a hint on the exotic nature of these states.

        Let us consider the experimental evidences on the large
apparent violation of the OZI rule trying to understand what
non-trivial physics is behind them.

\subsection{Antiproton \an~ at rest}

        The largest violation of the OZI rule was observed in
\ap~\an~ at rest. A question arises: why just this process is so
distinguished with respect to the OZI violation?

        The probable answer may connect with the specific of the
\ap~\an~ in the \ap -proton atom. The slow antiproton first
being captured on an orbit
of \ap -proton atom, preferentially with a large principal quantum number of
$n\sim 30$. Its further fate could be either a cascade to lower levels
or Stark mixing between the states with various angular momentum.
The relative probability of these two scenarios depends on the density of the
hydrogen. In gas at low pressure (of few millibars or less)
the main process is the cascade to lower levels and \an~ from the
P-levels with $n=2$. In liquid hydrogen, Stark mixing dominates and
the \ap~ annihilates from states with large principal number and
orbital angular momentum $L=0$, i.e. from the S-states.
The detailed discussion of these processes could be found
in \cite{Bat.89}, \cite{Gas.94}.

        So, in a first approximation, for \an~ in liquid hydrogen the
initial states are the S-levels.  In low density gas the \an~
mainly takes place from the P-levels. This circumstance strongly
facilitates the analysis of the reaction mechanism because the conservation of
C- and P-parities imposes strong restrictions on the allowed quantum
numbers of the initial state.

For instance, the reaction
$\bar{p}p \to \phi\pi^0$
is allowed only from two $\bar{p}p$
 initial states, $^3S_1$ and $^1P_1$.
The study of this reaction in liquid hydrogen gives us
information about the amplitude of this process from  the $^3S_1$
state, whereas in low density hydrogen gas we will obtain the information
on the $^1P_1$ amplitude. To reach the same goal for ordinary
$NN$ or $\bar{N}N$ interaction in flight we should use a polarized beam and
a polarized target and perform a partial amplitude analysis.

        Therefore, the important advantage of the \ap~\an~ at rest is
a possibility to obtain information about the amplitude of a process from
the initial state with fixed quantum numbers.

        The specific of the \ap~\an~ at rest is that the
measured values - annihilation frequencies or \an~ yields - do not
reflect directly
 the dynamics of the strong interaction but also depend on
the population of different atomic levels of the \p~atom.

The annihilation frequency of channel $Y$ is defined as the product of
the hadronic branching ratio $B$ by the fraction
$W(J^{PC},\rho)$ of annihilations
of the protonium atom from all the levels with given $J^{PC}$ at
the
given target density $\rho$.

For instance, the annihilation frequency of the $\phi\pi^0$ channel for
 the target of density $\rho$ is
\begin{eqnarray}
 Y(\bar{p}p\to \phi\pi^0,~\rho)
 & = & W(^3S_1,~\rho)\cdot B(\bar{p}p\to \phi\pi^0, ~^3S_1) + \nonumber \\
          & + & W(^1P_1,~\rho)\cdot B(\bar{p}p\to \phi\pi^0, ~^1P_1)
\label{Bat}
\end{eqnarray}
 In order to extract hadronic branching ratios,
 one has to know the annihilation fraction $W(J^{PC},\rho)$ for
each target
density $\rho$. Usually it was done from
the analysis of experimental data on different
\an~ reactions with using some information
from models of the cascade in \p~atom (see, cf., \cite{Batty},
\cite{Zoc.98}). It was assumed that the weights $W$ could be factorized as follows:
\begin{eqnarray}
 W(^3S_1,\rho) & = & 3/4 \cdot E(^3S_1,\rho) \cdot f_S(\rho)\\
 W(^1P_1,\rho) & = & 3/12 \cdot E(^1P_1,\rho) \cdot (1-f_S(\rho))
\end{eqnarray}
 Here $f_S(\rho)$ is the fraction of annihilations from the S-states,
 the numbers 3/4 and 3/12 are the statistical weights of the corresponding
 initial states and  $E(^3S_1,\rho)$ and $E(^1P_1,\rho)$ are the enhancement
factors which reflect deviations from the pure statistical population
of the levels. The enhancement factors $E(^3S_1)$ and $E(^1P_1)$, determined in
\cite{Batty}, turn out to be around 0.9-1 at all densities and $f_S(\rho)=0.87, 0.42, 0.20$
in liquid, gas target at NTP and 5 mbar, respectively. However, this set
of parameters is not unique and should be considered with some
warning ( see detailed discussion in \cite{Ben.00}).

\subsubsection{ $\bar{p} p \to \phi\gamma$}

The largest OZI rule violation is observed in the
$\bar{p} p \to \phi\gamma$
channel, where the Crystal Barrel collaboration has found~
\cite{Ams.98}, \cite{Ams.95} after phase space corrections:
\begin{equation}
R_{\gamma} = {B(\bar p p \rightarrow \phi \gamma )
\over B( \bar p p \rightarrow \omega \gamma)}
= (294\pm 97)\cdot 10^{-3},
\label{Rgamma}
\end{equation}
which is about 70 times larger than the
OZI prediction $R(\phi/\omega)=4.2\cdot10^{-3}$.

        The main experimental problem was the correct selection of the reaction
$\bar{p}p \to K_S K_L \gamma$ among a lot of events of
the $\bar{p}p \to K_S K_L \pi^0$ reaction.
If one of two $\gamma$ from $\pi^0$
decay had the energy less than the detection threshold of 10 MeV, then the two
channels were indistinguishable. The Monte Carlo simulation has
shown
that the probability of the feedthrough is small
$W(\phi \pi^0 \to \phi \gamma) = (0.52\pm0.02)\%$ \cite{Ams.95}.
However, the
yield of the reaction   $\bar{p} p \to \phi \pi^0$ is so large that
it gives $36\pm2$ background events in the $\phi \gamma$ data sample,
which after background subtraction  comprises $46\pm9$ events.
Therefore, the correction for the feedthrough is quite substantial,
at the level of 40\%.

        However,
the study of $\phi \gamma$ was done in two different channels:
$\bar{p}p \to K_S K_L \gamma$ and  $\bar{p}p \to K^+ K^- \gamma$.
The branching ratios extracted from the both reactions turned out to be similar.

        The reaction
$\bar p p \rightarrow \phi \gamma $
 is possible either from  the $^1S_0$ or from the
 $^3P_J$ states. The  measurements \cite{Ams.98}
were done in the liquid hydrogen, where the $^1S_0$ state is much more
probable than the $^3P_J$ states.
Thus, the apparent OZI rule violation was detected for
the $J^{PC}=0^{-+}$ initial state.

\subsubsection{ $\bar{p} p \to \phi\pi$}

Another very
large apparent violation of the OZI rule was found by the OBELIX and
Crystal Barrel collaborations in the
$ \bar{p} + p \to \phi + \pi$ channel.

For the ratio of the phase space corrected branching ratios
 the Crystal Barrel measurement \cite{Ams.98} in liquid hydrogen gives:

\begin{equation}
R_\pi={B(\bar p p \rightarrow \phi \pi )
\over B( \bar p p \rightarrow \omega \pi)}= (106\pm 12)\cdot 10^{-3}
\label{Rpi}
\end{equation}

It coincides with the ratio of the \an~ yields measured by the
OBELIX Collaboration
for annihilation in a liquid-hydrogen target~\cite{Don.98}:

\begin{equation}
R_\pi = (114\pm 10)\cdot 10^{-3}
\label{Rpitwo}
\end{equation}

The ratios (\ref{Rpi}) and (\ref{Rpitwo}) are
about a factor of 30 higher than the OZI rule prediction.

The OBELIX collaboration has performed a detailed investigation
of the $\phi \pi$ channel in different conditions.
First, the reaction
\begin{equation}
\bar{p} + p \to K^+ + K^- + \pi^0  \label{kkpi}
\end{equation}
was studied
for annihilation of stopped antiprotons in liquid hydrogen and in hydrogen
gas at normal temperature and pressure (NTP) and at
5 mbar pressure
\cite{Pra.98}.
It was accompanied by the measurements of the
$\bar{p} p \to \omega \pi^0$ channel for \an~in liquid hydrogen
and gas at NTP
\cite{Don.98}.

Second, the \an~ in gaseous deuterium at NTP
\begin{equation}
\bar{p} + d \to \pi^- + \phi(\omega) + p  \label{pd}
\end{equation}
was measured for two regions of spectator-proton momenta: $p<200$ MeV/c and
 $p>400$ MeV/c \cite{Abl.95}.

Third, the \an~of antineutrons
\begin{equation}
\bar{n} + p \to \pi^+ + \phi(\omega)   \label{an}
\end{equation}
 with momenta of 50-405 MeV/c was investigated \cite{Luc.93,Fil.98}.

The measurements of the reaction (\ref{kkpi}) at different hydrogen target densities
allow to establish an important feature of the apparent OZI rule violation
- its strong dependence on the quantum numbers of the initial state.

As we already mentioned, the fraction of annihilation from the
S-wave $f_S$ decreases
with decreasing of the target density. According to
 \cite{Batty}, $f_S$ is  87\% for \an~ in liquid hydrogen,
42\% and 20\% for gas at NTP and 5 mbar,
respectively.

Therefore, if the $\phi \pi$ channel occurs mainly from the
$^3S_1$  initial state, then the yield of the reaction should decrease
with decreasing of the hydrogen density. If it is dominated by
the $^1P_1$~initial state, the yield should grow up at a low pressure sample.

\begin{figure}
\epsfig{file=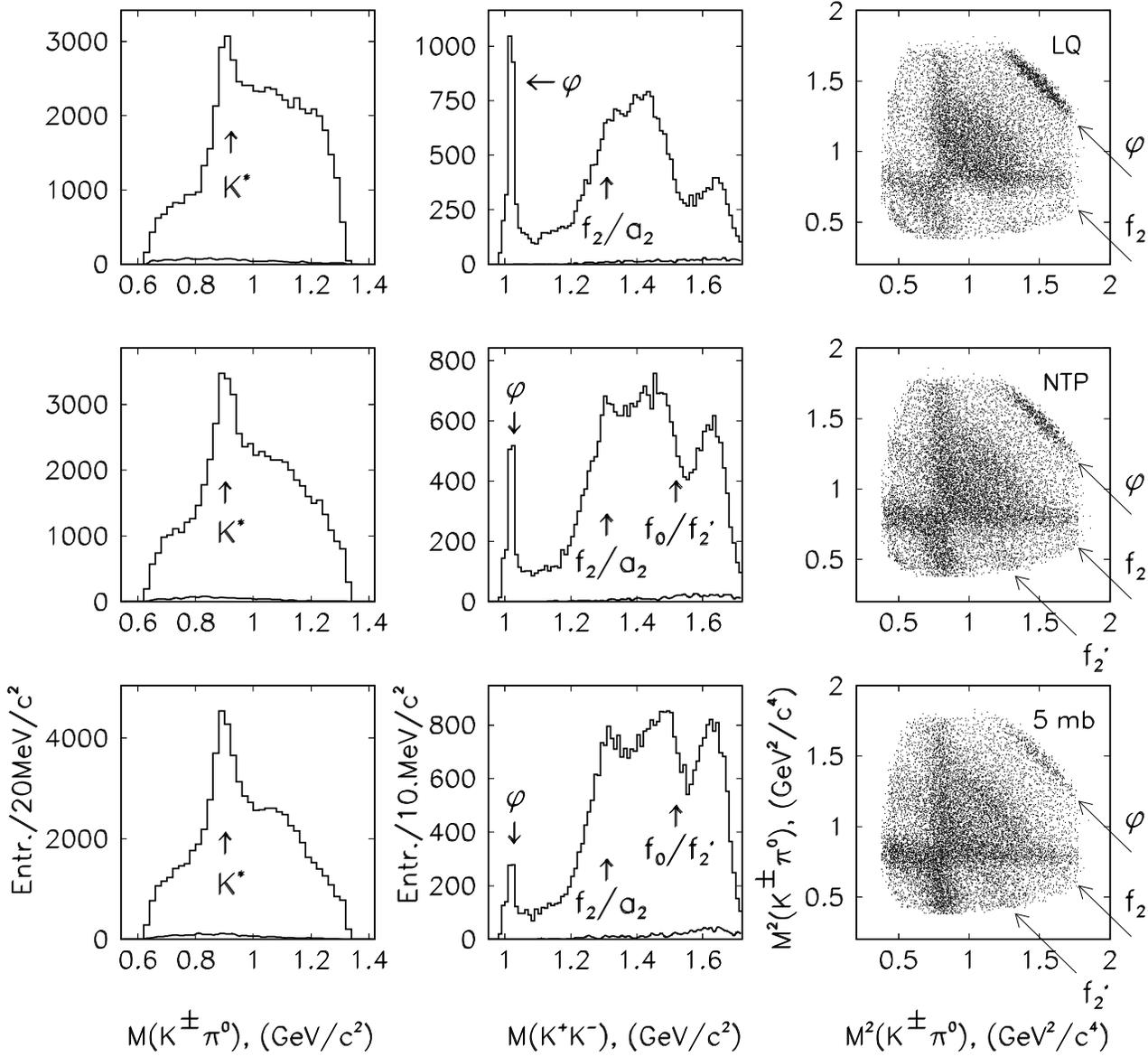,width=15.cm,height=15.cm,bbllx=1.5cm,bblly=0.cm,bburx=18.5cm,bbury=17.cm}
\caption{
Invariant mass distributions of the $K^+K^-$ (middle columns) and
$K^{\pm}\pi^0$ (left) systems
and
Dalitz plots (right) for
reaction
$\bar{p}p \to K^+K^-\pi^0$ at three different target densities:
for the liquid target (up line), for the gas target at NTP
(middle line) and for the gas target at 5 mb (bottom line)
\cite{Pra.98}.}
\label{fig:3pres}
\end{figure}

The measured
invariant mass distributions of the $K^+K^-$ and $K^{\pm}\pi^0$ systems
and the corresponding Dalitz plots for the reaction (\ref{kkpi})
at different hydrogen pressures are
shown in Fig.~\ref{fig:3pres}.

        One can see immediately three salient features of the spectra
in Fig.~\ref{fig:3pres}:\\
- the peak from the $\phi$ meson
reduces with decreasing of the density of the target;\\
- the peak from $K^*$ does {\it not} decrease with the density;\\
- the part of the $K^+K^-$ spectra with high invariant
mass ($M > 1.5$~GeV/$c^2$)
is more prominent in the low pressure data.

        All these features are important for the further analysis.
The enhancement in the  $K^+K^-$ spectrum around $1.5~ GeV/c^2$
at low pressure reflects the strong apparent violation of the OZI rule
for the tensor \ten~meson production. It will be discussed in
Section 3.1.4.
The density dependence of the $K^*$ meson production is important
for discriminating different models of the OZI violation
(see, Section 5.2). Here we will discuss the \f~meson production.

The dependence of the \f~ yield on the density clearly indicates
the dominance of the production from the $^3S_1$ state. The values of the
\an~ frequencies of \f~ and $K^*$ production
are given in the Table \ref{Reslt1}. Whereas
 total annihilation frequency of the $\bar{p}p \to K^+K^-\pi^0$ final state
 increases by about 50\% from the liquid to the low-pressure hydrogen
 target, the $\phi\pi^0$ yield in the same conditions
 decreases by more than 5 times.

\begin{figure}[p]
\epsfig{file=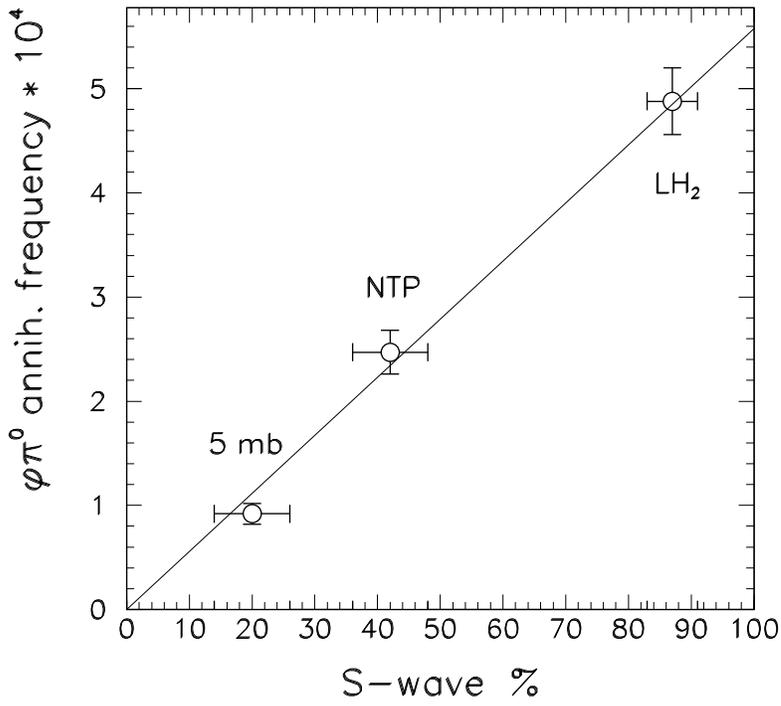,width=10.cm,height=10.cm,bbllx=1.cm,bblly=0.cm,bburx=16.cm,bbury=15.cm}
\caption{ Dependence of the $\bar{p} p \to \phi\pi^0$ annihilation frequency on
 the percentage of annihilation from the S-wave \cite{Pra.98}.
The values of the
 S-wave percentage at different target
 densities are from
 \protect\cite{Batty}~.}
\label{fig:phipi0sw}
\end{figure}

In Fig.~\ref{fig:phipi0sw} the dependence of the $\phi\pi^0$
annihilation frequency on the fraction of S-wave annihilations
is shown.
  One can see that the $\phi\pi^0$
 annihilation frequency linearly decreases with
percentage of the S-wave. It demonstrates the absence of the
contribution from the $^1P_1$
state.


Using the parameters of the  \p~ cascade
from \cite{Batty},
the branching ratios of the
$\bar{p}p \to \phi\pi^0$ from definite initial states
were determined \cite{Pra.98}:

\begin{eqnarray}
 B(\bar{p}p\to \phi\pi^0, ~^3S_1) & = & (7.57\pm0.62)\cdot 10^{-4}~,
\label{3s}\\
 B(\bar{p}p\to \phi\pi^0, ~^1P_1) & < & 0.5\cdot 10^{-4} ~~~~~~~~~~~~~~
\mbox{, with~95\%~CL}~ \label{1p}
\end{eqnarray}

The branching ratio of the $\phi\pi^0$ channel from
 the $^3S_1$ initial state is at least 15 times larger than that
 from the $^1P_1$ state. This demonstrates
 strong dependence of the $\phi\pi^0$ production on
the quantum numbers of the initial $\bar{p}p$ state.

 An indication of this selection rule was
 reported earlier in \cite{Rei.91, Abl.95b} though
on a scarce statistics.
 Thus in the OBELIX measurements \cite{Pra.98}
the number of $\phi$ events for 5 mbar data sample
 is $N_{\phi}=400\pm42$~, while the ASTERIX collaboration
 had only $N_{\phi}=4\pm4$
 for the data sample under similar conditions \cite{Rei.91}.

 It is important that
the observed tendency exists only for the \f~meson production,
the \om~ meson production from the P-wave is quite substantial.
That was observed in the measurements of the
$\bar{p}p \to \omega\pi^0$ channel at different hydrogen target densities
(the preliminary data is reported in \cite{Don.98}).
It turns out that the branching ratio of this reaction for the
\an~ from the $^1P_1$ state is not negligible.
The experimental angular distributions were fitted by the
sum of the angular distribution from the $^3S_1$ and $^1P_1$
initial states:

\begin{equation}
 W(\cos \Theta)= \alpha W_{^3S_1}(\cos \Theta) +
(1-\alpha ) W_{^1P_1}(\cos \Theta)
                   \label{ang }
\end{equation}
and the value of the $\alpha$ parameter was determined. This parameter
should be equal to 1 if the $\omega \pi$ final state as the $\phi\pi$
channel is dominated from the $^3S_1$ initial state. However, it
turns out that for \an~ in liquid hydrogen
$\alpha= 0.88 \pm 0.08$
and
$ \alpha= 0.69 \pm 0.10$
for \an~ in gas at NTP.

The non-negligible contribution of the $^1P_1$ state in the
$\omega \pi$ channel results in
different dependencies of the $\phi \pi$ and $\omega \pi$
\an~frequencies on the target density. Preliminary results \cite{Don.98}
for the
measurements of the ratio $R_{\pi} = Y(\phi\pi^0)/Y(\omega \pi^0)$ are:
\begin{eqnarray}
R_{\pi} = &(114\pm10)\cdot 10^{-3}~~~ &for~ LQ  \\
R_{\pi} = &( 83\pm10)\cdot 10^{-3}~~~ &for~ NTP \label{rntp}
\end{eqnarray}

        It is  also possible to obtain the values of the \f/\om~ ratio from
the  $^3S_1$ and  $^1P_1$ initial states. For that one should repeat
the same analysis for the $\omega \pi^0$ channel as it was done in \cite{Pra.98} for the $\phi \pi^0$.
Using $Y(\omega \pi^0)= (42.8\pm2.7)\cdot 10^{-4}$ for \an~ in
liquid hydrogen
and $Y(\omega \pi^0)= (29.6\pm2.7)\cdot 10^{-4}$ for \an~ in hydrogen gas
at NTP, one may arrive to the following important result:
\begin{eqnarray}
 R_{\pi}(\phi/\omega, ~^3S_1) & = & (120\pm 12)\cdot 10^{-3}~,
\label{r3s}\\
 R_{\pi}(\phi/\omega, ~^1P_1) & < & 7.2\cdot 10^{-3} ~~~~~~~~~~~~~~
 \mbox{, with~95\%~CL}~ \label{r1p}
\end{eqnarray}

        Therefore, a large apparent OZI violation occurs only for
\an~ from the spin--triplet S-wave initial state. When we
study the {\it same} reaction channel for spin--singlet P-wave state, the
disagreement with the OZI rule prediction magically disappears.

        It turns out that the large apparent OZI violation
could be switched off by changing the initial state
quantum numbers.

Investigations of the $\phi \pi$ channel for the \an~ in deuterium
$\bar{p} + d \to \pi^- + \phi(\omega) + p$
also confirm  the large apparent OZI rule violation
\cite{Abl.95}:
\begin{eqnarray}
 R_{\pi}(\phi \pi^-/\omega \pi^-) & = &
(133 \pm 26)\cdot 10^{-3}~,~~ p<200~MeV/c\\
 R_{\pi}(\phi \pi^-/\omega \pi^-) & = &
(113 \pm 30)\cdot 10^{-3}~,~~ p>400~MeV/c
\end{eqnarray}

        Here the basic reaction is $\bar{p} n \to \phi \pi^-$.
It was measured for two intervals of proton-spectator momenta:
$p<200$ MeV/c and $p>400$ MeV/c. The ratio $R_{\pi}$ is independent
of the momentum of the proton-spectator, indicating the relevant
dynamics to be connected with the basic $\bar{N}N$ interaction.

          From these results one may conclude that if
the apparent OZI-violation is due to excitation of some resonance, then
the resonance should have isospin  I=1 and $J^{PC}=1^{--}$.
We will discuss the resonance hypothesis in Section 5.2.

        The experiments with \an~ of antineutrons in flight
(\ref{an}) also have revealed the strong apparent OZI violation and
confirmed the difference between the $\phi \pi$ and $\omega \pi$ channels
\cite{Fil.98}.
The phase space corrected ratio of the branching ratios for \f~ and
\om~ production from S-wave is:
\begin{equation}
 R_{\pi}(\phi/\omega, ~^3S_1)  =  (112\pm 14)\cdot 10^{-3}
\end{equation}
that is in agreement with the measurements of Crystal Barrel and
OBELIX collaborations for the \ap~\an~ (\ref{Rpi}) and (\ref{r3s}).

The cross sections of the
$\bar{n}p \to \phi (\omega) \pi^+$ channels were measured
for antineutron momenta in the interval
50-405 MeV/c. It turns out that the
$\phi \pi^+$
cross section drops with energy, strictly following the decreasing of the S-wave. The
$\omega \pi^+$ cross section  decreases with energy not so rapidly.
A Dalitz-plot fit of the $\omega \pi^+$ final state
demands a significant P-wave contribution.
The branching ratio of the
$\omega \pi^+$ channel from the $^3S_1$ final state is
$B.R.(^3S_1)=(8.51\pm0.26\pm0.68)\cdot10^{-4}$, whereas
that of the
$^1P_1$ final state is only three times less,
$B.R.(^1P_1)=(3.11\pm0.10\pm0.25)\cdot10^{-4}$.
That is in sharp contrast with the hierarchy of the
same branching ratios for \an~
into $\bar{p}p \to \phi\pi$ final state (\ref{3s})-(\ref{1p}),
which differs by a factor of 15.

        It has been also found that the ratio
$R_{\pi} = \sigma(\phi\pi^+)/\sigma(\omega \pi^+)$  decreases with increasing
of the incoming antineutron momentum:
\begin{eqnarray}
 R_{\pi} = &
(100 \pm 17)\cdot 10^{-3}~,~~ & p=50-200~MeV/c\\
 R_{\pi} = &
(73.9 \pm 8.9)\cdot 10^{-3}~,~~& p=200-300~MeV/c\\
 R_{\pi} = &
(61.5 \pm 9.4)\cdot 10^{-3}~,~~& p=300-405~MeV/c
\end{eqnarray}

These results are important because they demonstrate how the large
apparent violation of OZI rule in the \an~ at rest smoothly
disappears with increasing the energy of the projectile and matches
with the results, shown in Table \ref{tpp} for \an~ in flight.

\subsubsection{ $\bar{p} p \to \phi\eta$}

        This channel was measured
by the OBELIX collaboration \cite{Nom.98} for
the $\bar pp$ annihilation at rest
in
liquid hydrogen, gas at NTP and at low pressure  of $5$ mbar.
The $\phi\eta$ final state has the same $J^{PC}$ as the
$\phi\pi^0$ final state. So, one may expect to see the same
selection rule as eqs.(\ref{3s})-(\ref{1p}) and suppose
that the \f~production in the low pressure sample will be
suppressed.
However, absolutely unexpectedly, the reverse trend is seen:
the yield of the $\bar{p}p \to \phi \eta$ channel grows
with decreasing of the target density.

\begin{figure}[p]
\psfull
\epsfig{file=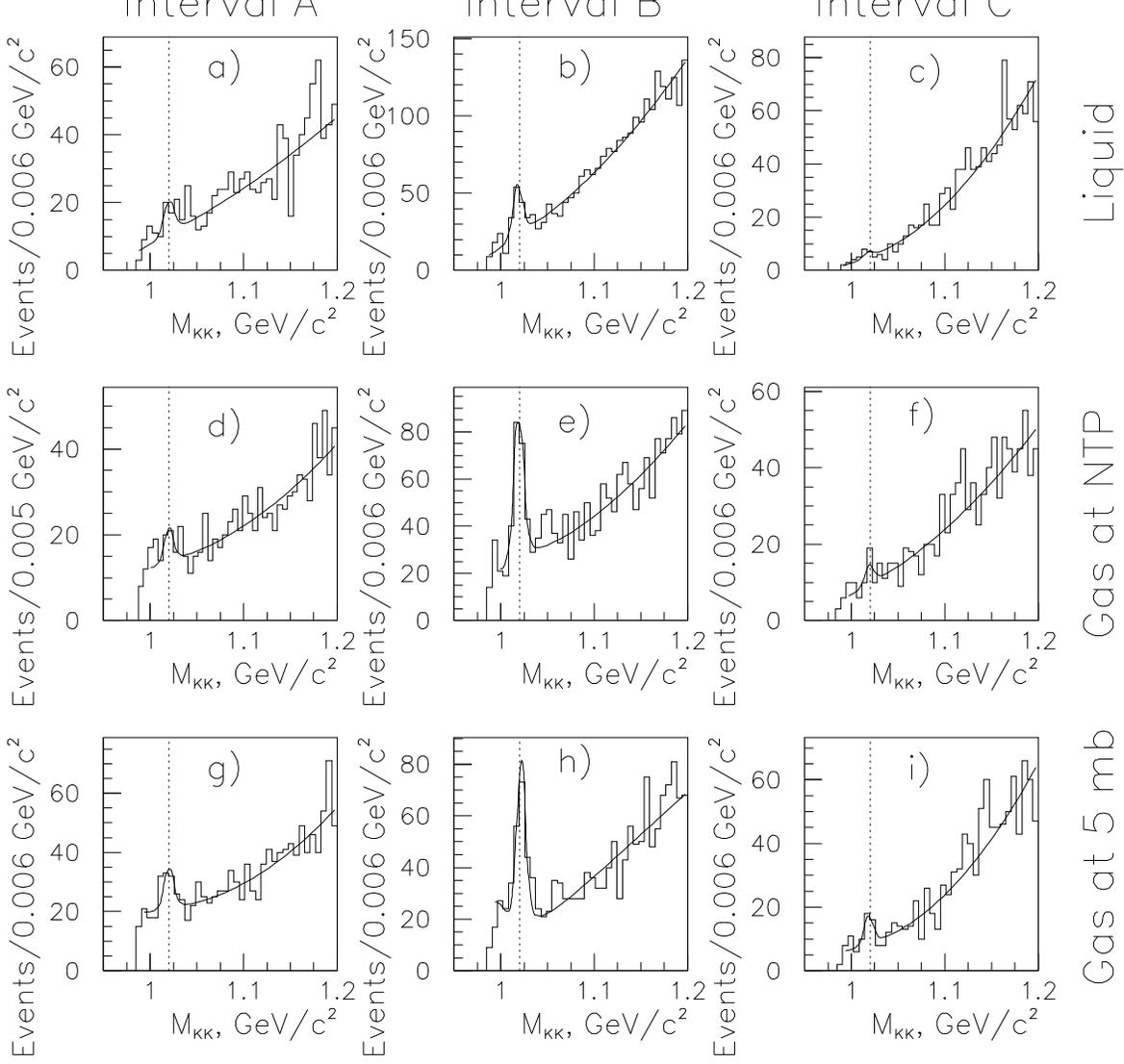,width=14.cm,height=14.cm,
bbllx=1.5cm,bblly=0.cm,bburx=18.5cm,bbury=17.cm}
\caption{The distributions on $M_{K^+K^-}$
for events of the reaction
$\bar{p}p \to K^+K^- X$
at three target densities: (a,b,c) for the liquid target,
(d,e,f) for the gas target at NTP and (g,h,i) for the gas
target at the 5 mbar pressure \cite{Nom.98}.
The column (a,d,g) corresponds to the interval of missing mass
below and  column (c,f,i) - above the eta mass.
 Dashed lines show
the $\phi$ meson mass.}
\label{fig:phieta}
\end{figure}

It is demonstrated in
 Fig. \ref{fig:phieta} where
the invariant mass distribution of two kaons $M_{K^+K^-}$
from the reaction $\bar{p}p \to K^+K^- X$ is shown.
In the middle part
of
Fig. \ref{fig:phieta}
the events with
the missing mass $0.26<M^2_{miss}<0.34~~GeV^2/c^4$
(centered around the mass of
$\eta$ meson, $m^2_{\eta}=0.3~~GeV^2/c^4$) are selected.
The left and right parts of
Fig. \ref{fig:phieta}
correspond to the $M_{K^+K^-}$ distributions for
the missing mass intervals below and above the $\eta$ mass:
$0.15<M^2_{miss}<0.23$ (left) and
$0.37<M^2_{miss}<0.45~~GeV^2/c^4$ (right).

One can see that the \f~peak in the $\eta$ missing mass interval is
growing up with the decreasing of the density.
Using the same parameters of the \p~ atom cascade for the evaluation
 of the branching ratios as in
\cite{Pra.98}, it is obtained \cite{Nom.98} that
\begin{eqnarray}
B(\bar{p}p \to\phi\eta, ^3S_1) & = &(0.76\pm 0.31)\cdot10^{-4} \label{r1} \\
B(\bar{p}p\to\phi\eta, ^1P_1) & = &(7.72\pm 1.65)\cdot10^{-4}  \label{r2}
\end{eqnarray}

        Therefore some dynamical selection rule is observed with a
trend
opposite to that for the $\phi\pi$ channel.

        The question arises: what about the OZI rule for the $\phi\eta$
production?

        The Crystal Barrel measurements of annihilation
  in liquid give
for the ratio of the phase space corrected branching ratios
\cite{Ams.98}

\begin{equation}
R_\eta={B(\bar p p \rightarrow \phi \eta )
\over B( \bar p p \rightarrow \omega \eta)}= (4.6\pm 1.3)\cdot 10^{-3}
\label{Reta}
\end{equation}
in a perfect agreement with the OZI--rule prediction for the
{\it vector} mesons (\ref{rfi}).

It will be interesting  to measure
the density dependence of the $\bar{p} p  \to \omega \eta$ channel.
The reasons for that are discussed in Section 5.5.

\subsubsection{ $\bar{p} p \to f_2'(1525) \pi^0$}

        The OBELIX measurements of the
$\bar{p} + p \to K^+ + K^- + \pi^0 $ channel
for annihilation of stopped antiprotons in
hydrogen targets of different density
\cite{Pra.98} provide for the first time the indication of the
apparent OZI rule violation for the tensor mesons.

        Discussing the Dalitz plots of this reaction at different
target densities shown in Fig.\ref{fig:3pres} we have already mentioned
that at the low pressure
the part of the $K^+K^-$ spectra with invariant
masses $M \sim 1.5$~GeV/$c^2$
is more prominent than in the \an~ in liquid.
The partial-wave analysis of the Dalitz plots \cite{Pra.98}
determines the yields of the tensor \s~
state - the \ten -meson and allows to compare them
with the S- and P-wave yields  of the $f_2(1270)$ meson, which consists
of light
quarks only.

 To avoid the problem of poor separation between the  $f_2(1270)$
 and  $a_2(1320)$ mesons,
the high statistics $\bar{p}p\to \pi^+\pi^-\pi^0$ channel was analysed
to determine the $\bar{p}p\to f_2(1270)\pi^0$ annihilation frequencies.
 In the $K^+K^-\pi^0$ final state there is strong interference
 between the $f_2(1270)$ and $a_2(1320)$ states,
 whereas in the $\pi^+\pi^-\pi^0$
 final state the $a_2(1320)$ contribution is absent.

Using the $f_2'$ yield from the analysis of the $K^+K^-\pi^0$ channel and
$f_2$ from the $\pi^+\pi^-\pi^0$ one, it was obtained \cite{Pra.98}
that
\begin{eqnarray}
 R(f_2'(1525)\pi^0/f_2(1270)\pi^0)
 & = &(47 \pm 14 ) \cdot 10^{-3} ~,~~~~\mbox{S-wave}  \label{f5} \\
 & = &(149 \pm 20 ) \cdot 10^{-3} ~,~~~\mbox{P-wave}  \label{f6}
\end{eqnarray}

    Remind the reader that the OZI--rule prediction for the
tensor mesons (\ref{rf2}) is
$R(f'_2 /f_2)=(3-16)\cdot10^{-3}$.

The result of (\ref{f5}) for the S-wave agrees with the Crystal Barrel
measurement \cite{Ams.98}
\begin{equation}
R(f_2'(1525)\pi^0/f_2(1270)\pi^0)= (26\pm10) \cdot 10^{-3}
\end{equation}
 for \an~
in liquid hydrogen, where the S-wave is dominant.
The excess of the \ten~production (\ref{f5}) observed in
the
S-wave is marginal within the experimental errors.
However, the
strong apparent violation of the OZI rule is seen
in the P-wave \an.
It means that for \an~in flight
this effect should increase with the energy of the incoming
antiproton.

\subsubsection{ $\bar{p} p \to \phi \pi \pi$}

 The OBELIX
collaboration has measured the reaction of the $\phi$ and
\om~production with two pions
\begin{equation}
\bar{p} + p \to \phi(\omega) + \pi^+ + \pi^-
\end{equation}
for
annihilation of stopped antiprotons in a gaseous and a
liquid hydrogen target \cite{Roz.96}.

The ratio $R_{\pi\pi} = Y(\phi\pi^+\pi^-) / Y(\omega\pi^+\pi^-)$
was determined for different invariant masses of the dipion system.
It turns out  that  without selection on the
dipion mass,
the ratio $R_{\pi\pi}$ is at the level of $(5-6) \cdot10^{-3}$, i.e.
in agreement with the prediction of the OZI rule.
However, at small dipion
masses
$300~ MeV < M_{\pi\pi} < 500~ MeV$
the degree of the OZI rule violation
increases up to $R_{\pi\pi} =(16-30) \cdot10^{-3}$.
That should be compared with the ratio $R_{\pi}
=(106\pm12) \cdot10^{-3}$  for the
$\phi\pi^0/\omega\pi^0$ channels for \an~in liquid
\cite{Ams.98}, which also proceeds from the same $^3S_1$
initial state.

\begin{figure}[p]
\epsfig{file=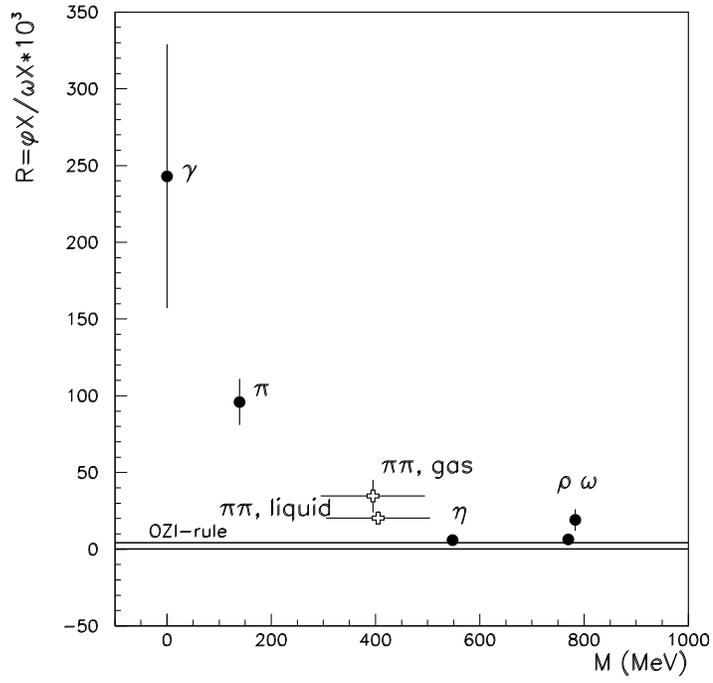,width=10.cm,height=10.cm,
bbllx=-3.cm,bblly=5.0cm,bburx=15.cm,bbury=23.cm}
\caption{The ratio $R=\phi X/\omega X $
corrected on phase space
for different reactions of $\bar pp$ annihilation at rest as
a function of the mass M of the system X (from \cite{Roz.96}).
Solid line shows the prediction of the
OZI rule.}
\label{momt}
\end{figure}

 In Fig. \ref{momt} the  values of the $R_{\pi\pi}$ corrected for the phase
 space difference
are compared with the results
of other measurements of binary reactions of
antiproton annihilation at rest.
The clear dependence of
the degree of the OZI rule violation on the mass  of the system
created
with $\phi$ is seen.
Namely, the degree of the OZI rule violation
increases with
decreasing of the mass of the system created with \f~.
For annihilation at rest  $\bar{N}N \to \phi X$ the decreasing of the
mass of X means an increase of the momentum transferred to the $\phi$.

The same effect was found in $\phi$ production in
$\pi^{\pm} N \to \phi N$ interaction
\cite{Coh.77} where the $d\sigma/dt$ distribution
of $\phi$  production at large $t$ differs significantly from the
one for $\omega$-meson, leading to the increase of $\phi/\omega$ ratio
at large $t$.
The direct measurements
of the t-dependence of the
differential cross sections of $\phi\pi$ and
$\omega\pi$ channels  in $\bar{p}p$ annihilation in flight should
clarify the problem.

It would be interesting to perform a systematic investigation to what
extent the
degree of the apparent OZI--rule violation depends on the momentum transfer.

\subsubsection{ $\bar{p} d \to \phi n$}

The largest momentum transfer in the $\phi$ production by stopped
antiproton annihilation is available in the so-called Pontecorvo reaction
\begin{equation}
  \bar{p} + d \to \phi + n \label{fin}
\end{equation}

        This is an example of a specific reaction of \ap~\an~
with only one meson in the final state
\begin{equation}
\bar{p} + d \to M +  N,
\end{equation}
where $M =\pi , \rho,\omega, \phi, \ldots$ is a meson
 and $N=p,n,\Delta,\Lambda,\ldots$ is a
baryon. They were first considered by B.Pontecorvo \cite{BMP} in 1956.
These processes are forbidden for \ap~\an~ in hydrogen but allowed for
\an~in deuterium.

The momentum transferred to the $\phi$
in the Pontecorvo reaction (\ref{fin})
with a stopped antiproton is
$q^2 = -0.762~GeV^2/c^2$, compared to $q^2 = -0.360~GeV^2/c^2$ for the
$\bar pp\rightarrow\phi\pi^0$ reaction.

        The OBELIX collaboration \cite{Cer.95} has measured
reaction
(\ref{fin})
in a deuterium gas target at NTP
and the Crystal Barrel collaboration
has measured the Pontecorvo reaction with $\omega n$ in the final
state \cite{Sch.93}.

 The kinematics of the Pontecorvo reaction
$\bar pd\rightarrow\phi n$ facilitates
its selection because it is the only two-body reaction of the
$\bar{p}d \to \phi X$
annihilation. The momentum of $\phi$ in
reaction (\ref{fin}) should be equal to 1.01 GeV/c for \an~ of \ap~
at rest.
Another quasi-two-body reaction is  $\bar pd\rightarrow\phi \pi^0 n$.
The yield of this
process is dominated by the
two-body annihilation on the
proton $\bar pp\rightarrow\phi \pi^0  $, whereas the neutron behaves
as a spectator with typical momenta $p \leq 200$ MeV/c.
The momentum of $\phi$ from this reaction is 650 MeV/c. It should be
spread by the Fermi motion of proton in deuteron and experimental resolution.

\begin{figure}[p]
\epsfig{file=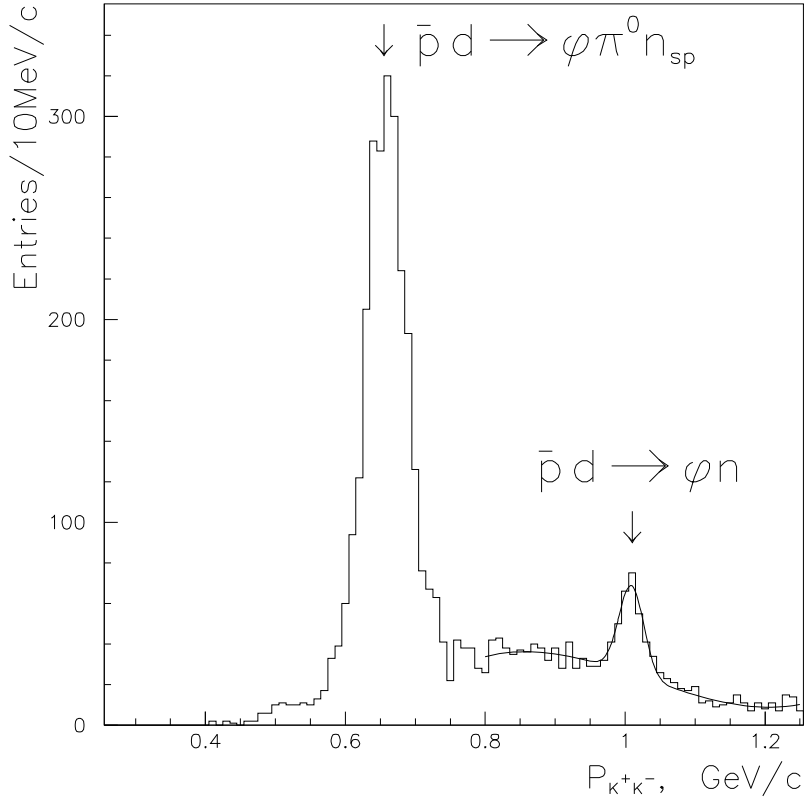,width=12cm}
\caption{The  total momentum
distribution of $K^+K^-$-system for the
events obeying the cut $|M_{K^+K^-}-M_{\phi}|< 10~MeV/c^2$
(from \cite{Cer.95}).}
\label{ponte}
\end{figure}

In Fig.\ref{ponte} the distribution of the events
on the total momentum of $K^+K^-$-system measured in \cite{Cer.95}
 is shown. To provide that
these two kaons are coming from \f~decay,
the
events with invariant mass around the mass of \f~ at
$|M_{K^+K^-}-M_{\phi}|< 10~MeV/c^2$ were chosen.
A clear peak  from the Pontecorvo reaction
is seen at $p_{K^+K^-}=1~ GeV/c$. A more prominent peak at
$p_{K^+K^-}=650~ MeV/c$ is mainly from
the reaction $\bar pd\rightarrow\phi \pi^0 n$.

The measurement
of the yield of (\ref{fin}) provides that
$Y(\bar pd\rightarrow\phi n) = (3.56 \pm0.20  ^{+0.2}_{-0.1}) \cdot10^{-6}$.
That should be compared with the yield of
the Pontecorvo reaction
for the $\omega$ production
${Y(\bar pd\rightarrow\omega n)=}$ ${(22.8\pm4.1)\cdot 10^{-6}}$ \cite{Sch.93}.
It turns out that the ratio $\phi/\omega$ is rather large:

\begin{equation}
R= Y(\bar{p} d\to \phi n)/Y(\bar{p} d\to\omega n) = (156 \pm 29)\cdot10^{-3} \label{rfin}
\end{equation}

It is even greater than the corresponding ratio $R_{\pi}$
from eq.(\ref{Rpi})
for the annihilation on a free nucleon
$\bar{p} p  \to  \phi \pi^0 $ in liquid hydrogen and
twice as large as
the ratio $R_{\pi}$ measured in hydrogen gas at NTP (\ref{rntp}).

        Therefore, there is a serious expectation
that the degree of the OZI rule violation depends on the momentum transfer.

\subsubsection{Summary of $\bar{p}p$ \an~ results}

The experiments at LEAR
with  antiproton \an~ have demonstrated the following distinctive
features:\\

1) Unusually strong deviation from the OZI--rule
predictions. In some reactions it exceed the OZI
estimation by a factor of 30-70.\\

2) This effect is not-universal for all \an~ channels of the \f~ production but
mystically occurs only in some of them.
For instance, no enhancement of the
\f~ production is observed for the $\phi \omega$ or $\phi\rho$
channels
($R(\phi \omega/\omega \omega) = (19\pm 7)\cdot 10^{-3}$,
$R(\phi \rho/\omega \rho) = (6.3\pm 1.6)\cdot 10^{-3}$
\cite{Sap.98}).
\\

3) There is a strong dependence of the OZI--rule violation on
the quantum numbers of the initial \p~state.
It was clearly demonstrated by the OBELIX collaborations results:
\begin{eqnarray}
 R_{\pi}(\phi/\omega, ~^3S_1) & = & (120\pm 12)\cdot 10^{-3}~,\\
 R_{\pi}(\phi/\omega, ~^1P_1) & < & 7.2\cdot 10^{-3} ~~~~~~~~~~~~~~
 \mbox{, with~95\%~CL}~
\end{eqnarray}

4) There is a serious indications
that the degree of the OZI rule violation depends on the
momentum transfer.\\

5) The apparent OZI-violation was found not only for the
\f~meson production but also for the tensor \s~ state -
\ten -meson. As in a case of \f~meson the apparent
OZI violation for tensor mesons
turns out to be extremely sensitive to the
quantum numbers of the initial state.

\subsection{Proton-proton and proton-deuteron interactions}

        The large apparent OZI violation was observed not only in
experiments with stopped antiprotons. Recent experiments with
\f~production in $pp$ and $pd$ interactions have also revealed
significant deviation from the OZI predictions.

\subsubsection{$pp \to p p \phi$}

 The DISTO collaboration \cite{Bal.98},\cite{Bal.00} has performed
the measurement of the \f~ and \om~
production
\begin{equation}
 p + p \to p + p +\phi (\omega) \label{ppfi}
\end{equation}
at the same proton energy of 2.85 GeV.

It was found \cite{Bal.00} that
\begin{equation}
R= \frac{\sigma(p p \to pp\phi)}{\sigma(p p \to pp\omega)} =
   (3.8 \pm 0.2 ^{+1.2}_{-0.9})\cdot10^{-3} \label{pp}
\end{equation}

At first glance this value seems to be in agreement with the OZI rule
prediction (\ref{rfi}). However, a correction on a different phase
space volume of the
$\phi pp$ and $\omega pp$ final states is needed. The incoming
proton energy of 2.85 GeV corresponds to 83 MeV above the \f~
production threshold and 320 MeV above the threshold
for the \om~ production. The corresponding
phase
spaces differ by a factor 14, on which one should multiply the ratio
(\ref{pp}).

Therefore, a substantial OZI violation has been observed
also in the proton-proton
interaction. One should mention that besides
the DISTO measurement, the  \f~production  in
$pp$ interaction  was investigated only in two other
experiments,  at 10 and 24 GeV/c \cite{Bal.77,Blo.75}.
The new experimental information
on the \f~and \om~production
near the threshold is badly
needed.

It is interesting that the DISTO collaboration
has observed also the
differences between the
angular distributions of the \f~ and \om~
mesons produced in (\ref{ppfi}).
The \f~meson angular distribution is flat, whereas
to fit the \om~ angular distribution, the first three even Legendre
polynomials are needed \cite{Bal.98}.

In general, the difference in the angular distributions of the
\f~ and \om~
mesons is an indication of different production
mechanisms. It contradicts the OZI rule postulate that the
\f~ could be formed in $pp$ interactions only via \om-\f~
mixing.

To demonstrate that, one should compare the \f~ and \om~
meson distributions at the same energy above the corresponding
thresholds. However, this condition was not fulfilled in the DISTO
measurements. So, in principle, it is possible that the  \om~ angular distribution
is more asymmetric simply  because it was measured at larger energy above the
threshold. Recently the \om~
production was measured at lower energy of 173 MeV above the
threshold \cite{Abd.01}. The measurement again reveals strong angular
anisotropy. That indicates that the difference in the \f~ and \om~
meson production is not due to kinematics but due to the difference in
the production mechanisms.

 Clearly future measurements should
avoid this disadvantage and take data at different incoming proton
energies but at the same energy above the threshold of the \f~ and \om~
production.

\subsubsection{$pp \to p p \phi$ diffractive production}

        Interesting results were obtained at the SPHINX spectrometer
at the Protvino U-70 accelerator \cite{Gol.97} in the study of diffractive
production of the \f~ and \om~mesons

\begin{equation}
 p + p \to p + [p V],~~ V=\phi,\omega  \label{dppfi}
\end{equation}
at 70 GeV proton energy.

        To fulfill the ``elasticity'' condition, to ensure the diffractive
character of the process,
the events were
selected in such a way that the total energy of the proton and the vector
meson decay products was within the 65-75 GeV interval.

        The simple ratio of the total cross sections of diffractively produced
\f~ and \om~ does not largely deviate from the OZI prediction of
(\ref{rfi}):

\begin{equation}
R= \frac{\sigma(p p \to p[p\phi])}{\sigma(p p \to p[p\omega])} =
   (15.5 \pm 0.5\pm3)\cdot10^{-3} \label{dpp}
\end{equation}

        However, the {\it shape} of the invariant mass spectra of the
$p\phi$ and $p\omega$ systems is absolutely different.

The authors  \cite{Gol.97}  argue that the simple ratio of the total
cross sections (\ref{dpp}) has little physical sense because a
significant contribution to the total cross section
of the $p\omega$ production comes from the kinematical region below
the threshold of the $p\phi$ production.
They prefer to compare either the differential cross sections of
the
vector mesons production with the same invariant mass of the $pV$ system,
or the differential cross sections in the same vector meson momentum interval.
 In both cases the (\f /\om) ratios
increase significantly till
$R(\phi/\omega) = (40-73) \cdot10^{-3}$.

It is interesting that this ratio of the differential cross sections
is practically independent of the  squared transverse momentum $p^2_T$
of the $pV$ system.

\subsubsection{$pd \to ^3He \phi$}

       The measurements of the $\phi$ and $\omega$ mesons production
 yields in
the reaction

\begin{equation}
p  + d \longrightarrow ^{3}He + \phi(\omega)      \label{hex}
\end{equation}
were performed at Saturne II
accelerator \cite{Wur.95}. The yield of $^{3}He$ was measured as a
function of the proton beam energy just near the thresholds of the \f~ and
\om~mesons production. The method exploits the fact that only
near the threshold the $^{3}He$ is formed at rest in the centre-of-mass
 frame and flies in
a very small cone around the beam direction covered by the
spectrometer.

A large deviation from the OZI rule
prediction $R(\phi/\omega) = 4.2\cdot10^{-3}$ was found:

\begin{equation}
R(\phi/\omega) = (80\pm3^{+10}_{-4})\cdot10^{-3}
\end{equation}

After corrections due to the \om-$^{3}He$ final state interactions
and different  \f~ and \om~thresholds
the apparent OZI violation is still large \cite{Wur.95}:
$ R(\phi/\omega) = (63\pm5^{+27}_{-8})\cdot10^{-3}$.

\subsection{Pion-proton interactions}

        In general, the agreement with the OZI rule predictions
is the best just for $\pi N$ scattering (see, discussion in
Sect.2.2). However, even in this case there is an experiment
\cite{Fer.97} which claims finding a large OZI violation (on factor 100).

        The authors of  \cite{Fer.97} have analysed the data of CERN
WA 56 experiment on the \f~ and \om~ production in a special kinematic regime:
\begin{equation}
 \pi^+ + p \to p_f + \phi(\omega) + \pi^+_s  \label{pifi}
\end{equation}
at beam momentum of 20 GeV/c (similar reactions for
$\pi^-p$ interaction were studied at 12 GeV/c).

Here $p_f$ stands for fast proton with lab momentum more than 10 GeV/c and
$\pi_s$ means slow pion, which was undetectable. Physically these conditions
correspond to the baryon exchange processes.

The measured  $\phi/\omega$ ratios  in the  reactions (\ref{pifi})
are enormous, ranging from
$R(\phi/\omega) = (430\pm 140)\cdot10^{-3}$ for $\phi\pi$ invariant
masses below 1.75 GeV till
$R(\phi/\omega) = (75\pm 34)\cdot10^{-3}$ for $\phi\pi$ invariant
masses  in the 2.75-3.25 GeV interval.

However, from the experimental point of view, there are some questions
concerning the measurement \cite{Fer.97}. The overall statistics is
scarce, comprises
$345\pm22$ events of \om -production and $247\pm22$ of \f -production.
In case of the \om~production there are two unseen pions in the reaction
(\ref{pifi}). In general, in such cases it is not possible to select
the reaction by the kinematical fit. The authors used additional experimental
information, but of course could not cure the situation completely.
In result, for instance, the invariant mass resolution in the \om~ peak is as large as
60 MeV.

So, these results need confirmation from the high statistics experiments with
reliable final state selection.

\subsection{ Production of the $\phi \phi$ system}

        Production of the $\phi \phi$ pair in the interactions of
non-strange mesons is a well known process which does not follow the
prescriptions of the OZI rule. Thus in \cite{Etk.82} it was found
that the yield of the reaction
$\pi^-p \to \phi \phi n$ at 22 GeV/c is significant.
It was not precisely compared with the corresponding yield
of the $\omega\omega$ channel but simply stated that the 100
observed events were too much for the OZI-forbidden reaction. The
effective mass of the $\phi \phi$ system  exhibits a large enhancement
around the 2.2-2.4 GeV region. The partial wave analysis has shown that
the largest contribution comes from the state with quantum numbers
of $J^PSL=2^+20$, where S is the total spin of two $\phi$ and L is
their orbital angular momentum. The
other significant contribution comes from $J^PSL=2^+22$ state,
again with S=2.

       The JETSET collaboration at LEAR has reported an
unusually high apparent violation
of the OZI rule in the \f \f~ final state   \cite{Eva.98} of
the \ap~\an:

\begin{equation}
  \bar{p} + p \to \phi + \phi \label{fifi}
\end{equation}

The measured cross section of
this reaction
turns out to be 2-4 $\mu$b  for the \ap~ momenta interval from
1.1 to 2.0 GeV/c.
One should compare it with the cross section of the reaction
$ \bar{p} + p \to \omega + \omega$, but this channel has not been
measured directly. The authors of  \cite{Eva.98} considered as a
reference point the cross section for the reaction
$ \bar{p} + p \to 2\pi^+ 2\pi^- 2\pi^0$, which is about 5 mb.
For the \om \om~ final state they put 10\%  of this value and assuming
standard mixing angle value obtained

\begin{equation}
 \sigma(\phi \phi) = \sigma(\omega \omega) \tan^4{(\Theta - \Theta_i)} \sim 10~nb
\label{fifi2}
\end{equation}

 It is by two orders of the magnitude less than the experimental value.

The spin-parity analysis of the JETSET data \cite{Ber.96} also shows
dominance of the $2^{++}$ states. It turns out that near the $\phi \phi$
threshold the largest contributions are from  $2^+$ states with the total spin of
the $\phi \phi$
system S=2.

The same trend was seen in the central production of the $\phi \phi$
system in $pp$ interactions \cite{Bar.98} at 450 GeV/c. A broad
enhancement was observed in the $\phi \phi$ effective mass spectrum
around 2.35 GeV. The angular analysis provides the best fit for the $2^+$ state with
S=2 and L=0.

 The comparison between the $\phi \phi$ production in $\pi^- p$ and
$K^-p$ interaction was done in \cite{Lan.96}. The
cross section of the OZI forbidden reaction
\begin{equation}
\pi^-p \to \phi \phi n \label{fi2}
\end{equation}
for pion momentum of 8 GeV/c was compared with the cross
section of the OZI allowed reaction
\begin{equation}
\pi^-p \to \phi K^+K^- n \label{fik}
\end{equation}
for the same kaon beam momentum.

It turns out that the ratio of these cross sections is

\begin{equation}
R_1=\frac{\sigma(\pi^-p \to \phi K^+K^- n)}{\sigma(\pi^-p \to \phi \phi
n)}=10.3 \pm 3.3
\end{equation}

Analogous processes for the $K^-p$ interaction are
\begin{equation}
K^-p \to \phi \phi \Lambda
\end{equation}
and
\begin{equation}
K^-p \to \phi K^+K^- \Lambda
\end{equation}
These reactions are OZI allowed and it was expected that the ratio
$R_2$
of their cross sections
\begin{equation}
R_2=\frac{\sigma(K^-p \to \phi K^+K^- \Lambda)}{\sigma(K^-p \to \phi \phi
\Lambda)}
\end{equation}
will be significantly different from the $R_1$. Namely, one
expects that $R_1 \gg R_2$. However, the experiment \cite{Lan.96}
found that $R_1 \simeq R_2$.

\vspace{0.5cm}
The violation of the OZI rule in the production of $\phi \phi$ system is so
drastic and clear
that the question of the explanation of these and other experimental results
discussed in this Section is mandatory. The proposal how to solve these
problems is discussed below.

\section{Polarized nucleon strangeness}

The role of the nucleon sea quarks is under extensive investigation
now. There are experimental indications that the
$\bar{s}s$ pairs in the proton
wave function are responsible for the number of non-trivial effects.

It was found that the magnitude of the strange quarks contribution
varies for different nucleon matrix elements. In \cite{Iof.90} an
explanation is giving, why the strange quarks contribution in the nucleon could be
at the same time small or large, depending on the considered matrix element.
The authors of \cite{Iof.90} connects the size of the nucleon strange matrix
elements with
the contribution from the QCD
nonperturbative effects. These nonperturbative effects should
increase the nucleon matrix elements $<p|\bar{s} O_n s|p>$ for
operators $O_n=\gamma_{\mu} \gamma_5, \gamma_5, I$.  Whereas for
$O_n=\gamma_{\mu}, \theta_{\mu \nu}=\gamma_{\mu}\partial_{\nu}+
\gamma_{\nu}\partial_{\mu}$ the nonperturbative effects are small
and an enhancement of the nucleon strange matrix elements is not
expected.

Indeed, the fraction of the nucleon
momentum carried by the strange quarks is
not large \cite{CCFR.95}, \cite{MRS.98}:
\begin{equation}
P_s= 4.\%~ at~Q^2=20~GeV^2
\end{equation}

The contribution of the strange quarks to the proton electric
form factor is also quite small. The HAPPEX Collaboration measurements
allow to extract the combinations of strange electric and magnetic form factors at
$Q^2=0.48~ (GeV/c)^2$ \cite{Ani.01}
\begin{equation}
G^s_E + 0.39 G^s_M= 0.025\pm0.020\pm0.014
\end{equation}
(the last error is related to
uncertainties in electromagnetic form factors).

The strange-quark contribution to the nucleon magnetic moment,
measured by the SAMPLE Collaboration \cite{Has.01}, is also small.
The contribution of strange quarks to the proton magnetic moment
is $-0.1\pm5.1\%$.

However
the contribution of the strange quarks in the nucleon mass may be
substantial.
Usually the strangeness content of the nucleon is parametrized as
the following ratio:
\begin{equation}
y = \frac{2<p|\bar{s}s|p>}{<p|\bar{u}u + \bar{d}d|p>} \label{y}
\end{equation}
The classical analysis of the $\pi N$ phase shifts \cite{Gas.91}
gives $y=0.2$ though with large uncertainties. The lattice
calculations give quite a large value of $y$: $y=0.36\pm0.03$
\cite{Don.96} and $y=0.59\pm0.13$ \cite{Gus.99}. The prediction of
heavy baryon chiral perturbation theory is
$y=0.20\pm0.12$\cite{Bor.99}. The calculation in framework of the
perturbative chiral model \cite{Lyu.00} gives somewhat small value
$y=0.076\pm0.012$. Also the cloudy bag model predicts a small value of
$y \sim 0.05$ \cite{Stu.97}.

 The summary of recent analysis
of the $\pi N$ experimental data for
evaluation of the nucleon $\sigma$-term was done in \cite{Mei.00}.
Main conclusion is quite unexpected, the
$\bar{s}s$ content of the proton turns out to be as large as $y=0.36-0.48$,
(see, also \cite{Pav.01}).

Moreover, during the past decade
the EMC \cite{EMC.88} and successor experiments
\cite{SMC.97}, \cite{E143}, \cite{E155}
with polarized lepton beams and
nucleon targets
gave indication that the \s\ pairs in the nucleon
are polarized:

\begin{equation}
\Delta s \equiv \int\limits_{0}^{1} dx[ s_{\uparrow}(x) -
s_{\downarrow}(x) + \bar s_{\uparrow}(x)-
\bar s_{\downarrow}(x)] = - 0.10\pm0.02.
\end{equation}
The minus sign means that the strange quarks and antiquarks are polarized
negatively with respect to the direction of the nucleon spin.

        Experiments on elastic neutrino scattering \cite{Gar.93} have also
provided an indication that the intrinsic nucleon strangeness is negatively
polarized though within large uncertainties.
It was obtained \cite{Gar.93} that $\Delta s = -0.15 \pm 0.07$.

        The analysis \cite{Ger.99} of the baryon magnetic moments
 also leads to the conclusion that in the proton the strange quarks are
 polarized and
$\Delta s = -0.19 \pm 0.05$.

The lattice QCD calculations  also indicate
the negative polarization of strange quarks in proton. The calculations
in the quenched approximation give  $\Delta s = -0.12 \pm 0.01$
\cite{Don.95} and $\Delta s = -0.109 \pm 0.030$ \cite{Fuk.95},
whereas the calculation in full lattice QCD \cite{Gus.99b} gives
$\Delta s = -0.12 \pm 0.07$.

        The negative polarization of the nucleon strange sea was
calculated within the framework of the SU(3) flavor chiral quark
model \cite{Che.01}. This model couples light quarks with octet of
pseudoscalar mesons by the requirement of the chiral symmetry.
 The values of the polarization within
$\Delta s = -(0.11-0.22)$ interval were predicted for different
model parameters.

     To explain the huge violation of the OZI rule in the annihilation
of stopped antiprotons and its
strong dependence on the spin of the initial state, which was discussed
in Sect. 3.1,
the model
based on a nucleon wave function containing negatively polarized
$s \bar s$ pairs
was proposed
\cite{Ell.95}, \cite{Ell.99}.

The model claims that the
observed OZI violation is only apparent because in these processes
the \s~meson is created via {\it connected} diagrams with participation
of intrinsic nucleon strange quarks.
The strong dependence on the
initial quantum numbers is due to polarization of the strange sea.
Let us discuss these assumptions in more details.

\subsection{Formation of \s~mesons}

        Let us consider the production of \s~ strangeonia in
$NN$ or $\bar{N}N$ interactions assuming
that the nucleon wave function
contains an admixture of $\bar{s}s$ pairs which are polarized negatively
with respect to the direction of the nucleon spin.

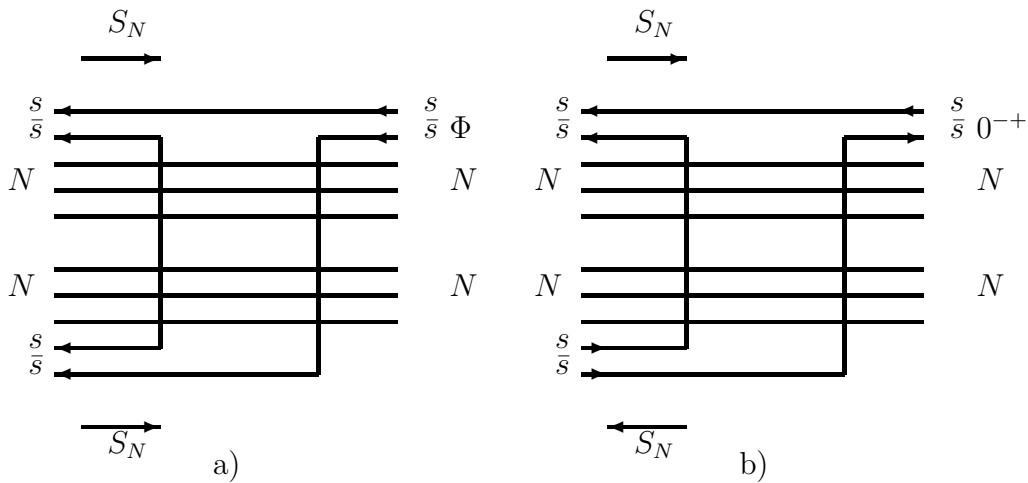
\begin{figure}[htb]
\setlength {\unitlength} {0.7mm} \thicklines
\linethickness{0.5mm}
\begin{picture}(180,70)(0,10)
\put(60,20){\vector(-1,0){50}}

\put(30,25){\vector(-1,0){20}}

\put(10,30){\line(1,0){65}}

\put(10,35){\line(1,0){65}}

\put(10,40){\line(1,0){65}}


\put(70,70){\vector(-1,0){60}}

\put(30,65){\vector(-1,0){20}}

\put(10,60){\line(1,0){65}}

\put(10,55){\line(1,0){65}}

\put(10,50){\line(1,0){65}}


\put(30,65){\line(0,-1){40}}

\put(60,20){\line(0,1){45}}


\put(75,70){\vector(-1,0){5}}

\put(75,65){\vector(-1,0){5}}

\put(60,65){\line(1,0){10}}

\put(75,35){\line(-1,0){20}}

\put(75,30){\line(-1,0){20}}


\put(40,1){a)}

\put(5,20){$\bar{s}$}

\put(5,25){$s$}

\put(5,65){$\bar{s}$}

\put(5,70){$s$}

\put(80,65){$\bar{s}$}

\put(80,70){$s$}

\put(85,65){$\Phi$}

\put(1,35){$N$}

\put(1,55){$N$}

\put(85,35){$N$}

\put(85,55){$N$}

\put(20,85){$S_{N}$}

\put(20,5){$S_{N}$}


\linethickness{0.5mm}

\put(15,10){\vector(1,0){15}}

\put(15,80){\vector(1,0){15}}



\put(110,20){\vector(1,0){5}}

\put(160,20){\line(-1,0){45}}

\put(110,25){\vector(1,0){5}}

\put(130,25){\line(-1,0){15}}

\put(110,30){\line(1,0){65}}

\put(110,35){\line(1,0){65}}

\put(110,40){\line(1,0){65}}


\put(170,70){\vector(-1,0){60}}

\put(130,65){\vector(-1,0){20}}

\put(110,60){\line(1,0){65}}

\put(110,55){\line(1,0){65}}

\put(110,50){\line(1,0){65}}


\put(130,65){\line(0,-1){40}}

\put(160,20){\line(0,1){45}}


\put(175,70){\vector(-1,0){5}}

\put(170,65){\vector(1,0){5}}

\put(160,65){\line(1,0){10}}

\put(175,35){\line(-1,0){20}}

\put(175,30){\line(-1,0){20}}


\put(140,1){b)}

\put(105,20){$\bar{s}$}

\put(105,25){$s$}

\put(105,65){$\bar{s}$}

\put(105,70){$s$}

\put(180,65){$\bar{s}$}

\put(180,70){$s$}

\put(185,65){$0^{-+}$}

\put(101,35){$N$}

\put(101,55){$N$}
\put(185,35){$N$}
\put(185,55){$N$}
\put(120,85){$S_{N}$}
\put(120,5){$S_{N}$}
\linethickness{0.5mm}
\put(130,10){\vector(-1,0){15}}
\put(115,80){\vector(1,0){15}}
\end{picture}
\vspace*{1cm}
\caption{Production of the \s~ mesons in $NN$ interaction
from the spin-triplet (a) and spin-singlet (b) states.
The arrows
show the direction of spins of the nucleons and strange quarks.}
\label{fig:8}
\end{figure}

        Due to the interaction it is possible that these pairs could
be either shaken-out from the nucleon or strange quarks from different
nucleons
could participate in some rearrangement process similar to the one shown in
Fig. \ref{fig:8}. Let us assume further that the quantum numbers of the
\s~pair is $J^{PC}=0^{++}$ (later we will explain this choice).

        Then the shake-out of such pairs will not create \f~
or tensor \ten~meson, but
a scalar strangeonium. The \s~ systems with other quantum numbers
(like \f~ or \ten)  should be produced due to a process where
strange quarks from {\it both} nucleons are participating.

 The examples of
these rearrangement diagrams are shown in Fig.~\ref{fig:8}.
  If the nucleon spins are parallel (Fig.~\ref{fig:8}a),
then the spins of
the $\bar{s}$ and $s$ quarks in both nucleons are also parallel.
If the polarization of the strange quarks does not change during the
interaction, then the $\bar{s}$ and $s$ quarks could keep parallel
spins in the final state. The total spin of \s~ quarks will be $S=1$ and
if their relative orbital momentum is $L=0$, it means that the
strangeonium has the \f~ quantum numbers, if $L=1$, it will correspond to
the creation of tensor strangeonium, $f'_2(1525)$.

        If the initial $NN$ state is a spin-singlet, the spins of
strange quarks in different nucleons are antiparallel and
the rearrangement diagrams like that in Fig.\ref{fig:8}b may lead
to the preferential formation of the \s~ system with total spin $S=0$. It means
that for $L=0$ one should expect additional production of strangeonia with the pseudoscalar quantum numbers
$0^{-+}$.

Therefore, the rearrangement of the \s~pairs always takes place, but
it does not mean that always the \s~pairs are produced as a $\phi$ meson.
The predictions of the polarized strangeness model for \ap~\an~
are quite definite:

\begin{itemize}
\item the \f~ should be produced mainly from the $^3S_1$ state
\item the \ten~  should be produced mainly from the $^3P_J$ states
\item the spin-singlet initial states favour the formation of
pseudoscalar strangeonia.
\end{itemize}

It explains why the $\phi/\omega$ ratio is not the same in different
channels of \ap~\an~ at rest. It is simply due to different initial states.

The polarized strangeness model postulates that the OZI rule itself is valid and
its observed violation is only apparent. It means that one could not describe
the \f~ production only via disconnected diagrams. The \f~meson
may be produced also in the processes described by the connected quark diagrams
if the
nucleon structure is complicated and allows the presence of the
polarized strange sea.

         It is important to note that these rules should be
preserved for antiproton-proton annihilation, as well as for nucleon-nucleon
interaction. The natural question is why the
especially large OZI violation was observed in the \ap~\an~ at rest whereas
the same reaction in flight
 exhibits no deviation from the OZI predictions.

        The answer is clear from the general feature of the \ap~\an~
at rest which was discussed in Sect. 3.1. The conservation of C- and
P-parities selects only few initial states with definite values of total
spin and angular momentum.
In this sense \an~in the \p~ atom is an analog of polarized antinucleon interaction with
polarized nucleon target. In these conditions, when the initial state
quantum numbers are fixed (or strongly limited),
 we could obtain more detailed
information about the interaction amplitude.

\subsection{Quantum numbers of the nucleon \s~pairs}

        There are different possibilities for the quantum
numbers of the \s~ component in the nucleon wave function.
It may have, for instance,
pseudoscalar quantum numbers $J^{PC} = 0^{-+}$ or vector $J^{PC} = 1^{--}$ ones.
Then the relative angular momentum $j$ between
the \s~ and $uud$ clusters with $J^P = 1/2^+$ should be $j=1$. However, it
is also possible that
the \s~pair has quantum numbers of the vacuum $J^{PC} = 0^{++}$,
then $j=0$ to provide quantum numbers of proton. It is up to the experiment
to determine which of these possibilities are realized in nature.

        One could see that the \s~ could be stored in the nucleon with the
quantum numbers of $\eta$ and $\phi$
if the relative angular momentum
between the \s~ and the $uud$ clusters is $j=1$. But if
$j=0$, then the quantum numbers of
\s~ pair may be
different, including the vacuum quantum
numbers $J^{PC}=0^{++}$.
Predictions of the model will depend drastically
on the assumption about the \s~ quantum numbers.
Thus, the assumption that the \s~ pair has
quantum numbers of \f-meson leads to serious problems.
 In this case one might expect some additional
\f~ production due to the strangeness, stored in the nucleon.
This
quasi-\f~ pair could be easily shaken-out from the nucleon.
Then it is not clear how to explain the strong dependence of the
\f~ yield on quantum numbers of {\it both} nucleons, discussed in
Sect.3.1.2.

Moreover, the shake-out of the \f~ stored in the nucleon
should lead to an apparent violation  of the
OZI rule
in {\it all}
reactions of the \f~ production.

This is ruled out by the
experiment, which showed that the OZI--rule violation is not
a universal trend of
all channels of \f~production in \p~
\an. In general the OZI--rule is fulfilled within 10 \% accuracy, but
there are few cases of the strong (on factor 30-70)
violation of the OZI--rule.

        Similar arguments were provided in \cite{Dov.90}, where
it was demonstrated that the experimental
data on the production of $\eta$ and $\eta'$ mesons exclude the $0^{-+}$
quantum numbers for the \s~ admixture in the nucleon wave function.

        In \cite{Alb.95} it was argued that the strange nucleon
sea may be negatively polarized due to the interaction of the light
valence quarks with the QCD vacuum. Due to the chiral dynamics
the interaction between quarks and antiquarks is most strong in
the pseudoscalar $J^{PC}=0^{-+}$ sector. This strong attraction
in the spin--singlet pseudoscalar channel between light valence
quark from the proton wave function and a strange antiquark from
the QCD vacuum will result in the spin of the strange antiquark
which
will be aligned opposite to the spin of the light quark (and, finally,
opposite to the proton spin). As strange antiquark comes from the
vacuum, the corresponding strange quark to preserve the vacuum
quantum numbers $J^{PC}=0^{++}$ should also be aligned opposite to
the nucleon spin.

From the QCD sum rules analysis \cite{Iof.81},\cite{Rei.84}
it is known that the condensate of the strange
quarks in the vacuum is not small and comparable with the condensate
of the light quarks:

\begin{equation}
< 0\left|\bar{s}s \right|0> = (0.8\pm0.1)< 0\left|\bar{q}q \right|0>,~~
q=(u,d)
\end{equation}

Thus, the density of $\bar{s}s$ pairs in the QCD vacuum is quite high and
one may expect that the effects of the polarized strange quarks in
the nucleon will
be also non-negligible.

        Therefore, we arrive to the picture of the negatively polarized
\s~ pair with the vacuum quantum numbers $^3P_0$.
These strange
quarks should not be considered like constituent quarks formed
some five quarks configuration of the nucleon. Rather they are
included in the components of a constituent quark.
It is important to stress that the \s~ pair with the $^3P_0$
quantum numbers  itself is not polarized  being a scalar.
That is a chiral non--perturbative interaction which selects only
one projection of the  total spin of the \s~ pair on the direction of
the nucleon spin.

\subsection{Shake-out of the intrinsic \s~pairs}

        Since the quantum numbers of the \s~pair in nucleon
are fixed to be $^3P_0$, the straightforward prediction is that one should see
the shake-out of this state in the nucleon-nucleon or \ap-proton
interactions. If the strange scalar strangeonium is $f_0(980)$
meson \cite{PDG},
then the shake-out of the intrinsic strangeness should lead to
some enhancement of the $f_0(980)$ production. The rearrangement diagrams like those in
Fig.~\ref{fig:8} should in general lead to
increasing of the $f_0(980)$ yield for \an~ from the
P-wave. This effect is the same as observed for the production
of tensor strangeonium. However  one should observe
increasing  of the ratio of $f_0(980)$ yield to the yield of light quark
state $\sigma(400-1200)$ for \an~ from the P-wave.
It is hard to determine this ratio due to a large width of
the $\sigma$ meson.

        However the shake-out of $^3P_0$ state with its subsequent
decay into kaons should lead to some quite peculiar
effects. In \cite{Ell.99} it was stressed that  shake--out of the
negatively-polarized
\s\ pair from the $^3S_1$ initial state should lead to the enrichment of
charged $K^+K^-$ pairs over the $K^0 \bar{K}^0$ ones and
neutral $K^{*0}\bar{K}^{*0}$ over $K^{*+} K^{*-}$.

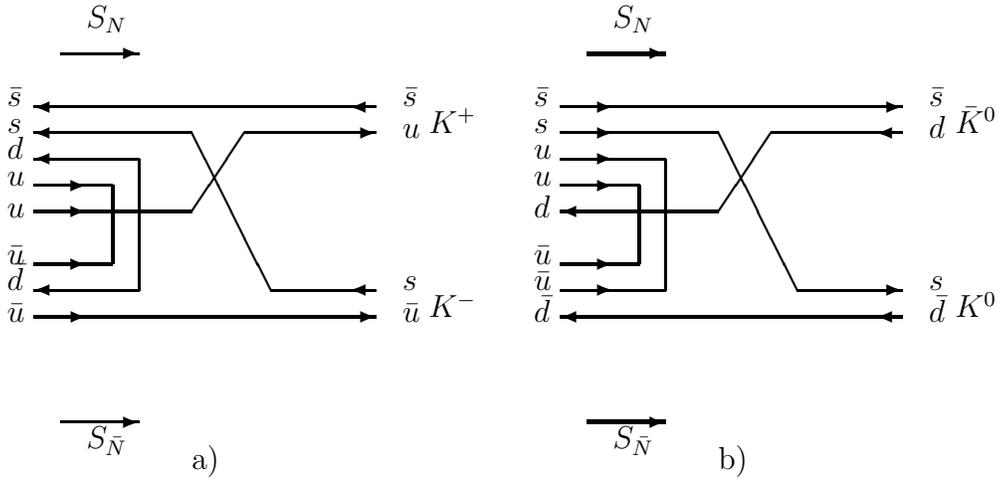
\begin{figure}[htb]


\linethickness{0.5mm}
\setlength {\unitlength} {0.7mm} \thicklines
\begin{picture}(180,70)(0,10)


\put(10,30){\vector(1,0){10}}
\put(20,30){\line(1,0){50}}
\put(15,35){\vector(-1,0){5}}
\put(15,35){\line(1,0){15}}
\put(10,40){\vector(1,0){10}}
\put(20,40){\line(1,0){5}}
\put(70,70){\vector(-1,0){60}}
\put(40,65){\vector(-1,0){30}}
\put(15,60){\vector(-1,0){5}}
\put(15,60){\line(1,0){15}}
\put(10,55){\vector(1,0){10}}
\put(20,55){\line(1,0){5}}
\put(10,50){\vector(1,0){10}}
\put(20,50){\line(1,0){20}}
\put(30,60){\line(0,-1){25}}
\put(25,55){\line(0,-1){15}}
\put(40,65){\line(1,-2){15}}
\put(40,50){\line(2,3){10}}
\put(75,70){\vector(-1,0){5}}
\put(70,65){\vector(1,0){5}}
\put(70,65){\line(-1,0){20}}
\put(60,65){\line(1,0){10}}
\put(75,35){\vector(-1,0){5}}
\put(70,35){\line(-1,0){15}}
\put(70,30){\vector(1,0){5}}
\put(60,30){\line(1,0){15}}
\put(40,1){a)}
\put(5,70){$\bar{s}$}
\put(5,65){$s$}
\put(5,60){$d$}
\put(5,55){$u$}
\put(5,49){$u$}
\put(5,40){$\bar{u}$}
\put(5,35){$\bar{d}$}
\put(5,29){$\bar{u}$}

\put(80,70){$\bar{s}$}
\put(80,64){$u$}
\put(80,29){$\bar{u}$}
\put(80,35){$s$}
\put(85,65){$K^{+}$}
\put(85,30){$K^{-}$}
\put(20,85){$S_{N}$}
\put(20,5){$S_{\bar{N}}$}

\put(15,10){\vector(1,0){15}}
\put(15,80){\vector(1,0){15}}



\put(115,30){\vector(-1,0){5}}
\put(115,30){\line(1,0){60}}

\put(110,35){\vector(1,0){10}}
\put(120,35){\line(1,0){10}}

\put(110,40){\vector(1,0){10}}
\put(120,40){\line(1,0){5}}

\put(110,70){\vector(1,0){10}}
\put(120,70){\line(1,0){50}}

\put(110,65){\vector(1,0){10}}
\put(120,65){\line(1,0){20}}

\put(110,60){\vector(1,0){10}}
\put(120,60){\line(1,0){10}}

\put(110,55){\vector(1,0){10}}
\put(120,55){\line(1,0){5}}

\put(115,50){\vector(-1,0){5}}
\put(115,50){\line(1,0){25}}

\put(130,60){\line(0,-1){25}}
\put(125,55){\line(0,-1){15}}
\put(140,65){\line(1,-2){15}}
\put(140,50){\line(2,3){10}}
\put(170,70){\vector(1,0){5}}

\put(175,65){\vector(-1,0){5}}

\put(170,65){\line(-1,0){20}}
\put(160,65){\line(1,0){10}}

\put(170,35){\vector(1,0){5}}
\put(170,35){\line(-1,0){15}}
\put(175,30){\vector(-1,0){5}}
\put(170,30){\line(-1,0){15}}
\put(140,1){b)}
\put(105,70){$\bar{s}$}
\put(105,65){$s$}
\put(105,60){$u$}
\put(105,55){$u$}
\put(105,49){$d$}
\put(105,40){$\bar{u}$}
\put(105,35){$\bar{u}$}
\put(105,29){$\bar{d}$}

\put(180,70){$\bar{s}$}
\put(180,64){$d$}
\put(180,29){$\bar{d}$}
\put(180,35){$s$}
\put(185,65){$\bar{K}^0$}
\put(185,30){$K^{0}$}
\put(120,85){$S_{N}$}
\put(120,5){$S_{\bar{N}}$}
\linethickness{0.5mm}
\put(115,10){\vector(1,0){15}}
\put(115,80){\vector(1,0){15}}

\end{picture}
\vspace*{1cm}
\caption{\it Production of $K\bar{K}$ due to shake-out  of a polarized
\s\ pair in the proton wave function
in $\bar{p}p$ annihilation from the initial $^3S_1$ state,
for (a) negative and (b) positive polarization of the \s\ pair.
The arrows show the directions of the spins of the nucleons and
quarks.}
\label{kk}
\end{figure}

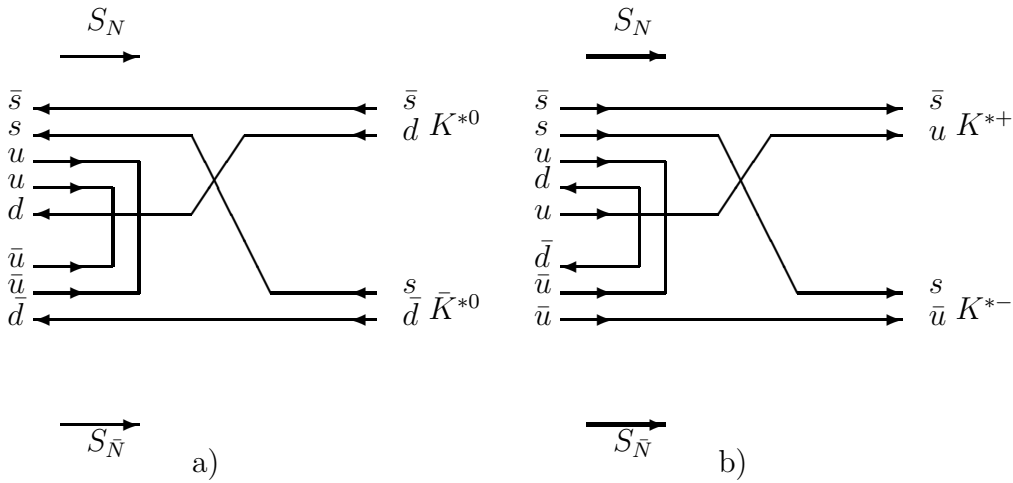
\begin{figure}[htb]


\linethickness{0.5mm}
\setlength {\unitlength} {0.7mm} \thicklines
\begin{picture}(180,70)(0,10)


\put(15,30){\vector(-1,0){5}}
\put(15,30){\line(1,0){60}}
\put(10,35){\vector(1,0){10}}
\put(20,35){\line(1,0){10}}
\put(10,40){\vector(1,0){10}}
\put(20,40){\line(1,0){5}}
\put(70,70){\vector(-1,0){60}}
\put(40,65){\vector(-1,0){30}}
\put(10,60){\vector(1,0){10}}
\put(20,60){\line(1,0){10}}
\put(10,55){\vector(1,0){10}}
\put(20,55){\line(1,0){5}}
\put(20,50){\vector(-1,0){10}}
\put(20,50){\line(1,0){20}}
\put(30,60){\line(0,-1){25}}
\put(25,55){\line(0,-1){15}}
\put(40,65){\line(1,-2){15}}
\put(40,50){\line(2,3){10}}
\put(75,70){\vector(-1,0){5}}
\put(75,65){\vector(-1,0){5}}
\put(70,65){\line(-1,0){20}}
\put(60,65){\line(1,0){10}}
\put(75,35){\vector(-1,0){5}}
\put(70,35){\line(-1,0){15}}
\put(75,30){\vector(-1,0){5}}
\put(70,30){\line(-1,0){15}}
\put(40,1){a)}
\put(5,70){$\bar{s}$}
\put(5,65){$s$}
\put(5,60){$u$}
\put(5,55){$u$}
\put(5,49){$d$}
\put(5,40){$\bar{u}$}
\put(5,35){$\bar{u}$}
\put(5,29){$\bar{d}$}

\put(80,70){$\bar{s}$}
\put(80,64){$d$}
\put(80,29){$\bar{d}$}
\put(80,35){$s$}
\put(85,65){$K^{*0}$}
\put(85,30){$\bar{K}^{*0}$}
\put(20,85){$S_{N}$}
\put(20,5){$S_{\bar{N}}$}

\put(15,10){\vector(1,0){15}}
\put(15,80){\vector(1,0){15}}



\put(110,30){\vector(1,0){10}}
\put(120,30){\line(1,0){55}}

\put(110,35){\vector(1,0){10}}
\put(120,35){\line(1,0){10}}

\put(120,40){\vector(-1,0){10}}
\put(120,40){\line(1,0){5}}

\put(110,70){\vector(1,0){10}}
\put(120,70){\line(1,0){50}}

\put(110,65){\vector(1,0){10}}
\put(120,65){\line(1,0){20}}

\put(110,60){\vector(1,0){10}}
\put(120,60){\line(1,0){10}}

\put(120,55){\vector(-1,0){10}}
\put(120,55){\line(1,0){5}}

\put(110,50){\vector(1,0){10}}
\put(120,50){\line(1,0){20}}

\put(130,60){\line(0,-1){25}}
\put(125,55){\line(0,-1){15}}
\put(140,65){\line(1,-2){15}}
\put(140,50){\line(2,3){10}}
\put(170,70){\vector(1,0){5}}

\put(170,65){\vector(1,0){5}}

\put(170,65){\line(-1,0){20}}
\put(160,65){\line(1,0){10}}

\put(170,35){\vector(1,0){5}}
\put(170,35){\line(-1,0){15}}
\put(170,30){\vector(1,0){5}}
\put(170,30){\line(-1,0){15}}
\put(140,1){b)}
\put(105,70){$\bar{s}$}
\put(105,65){$s$}
\put(105,60){$u$}
\put(105,55){$d$}
\put(105,49){$u$}
\put(105,40){$\bar{d}$}
\put(105,35){$\bar{u}$}
\put(105,29){$\bar{u}$}

\put(180,70){$\bar{s}$}
\put(180,64){$u$}
\put(180,29){$\bar{u}$}
\put(180,35){$s$}
\put(185,65){$K^{*+}$}
\put(185,30){$K^{*-}$}
\put(120,85){$S_{N}$}
\put(120,5){$S_{\bar{N}}$}
\linethickness{0.5mm}
\put(115,10){\vector(1,0){15}}
\put(115,80){\vector(1,0){15}}

\end{picture}
\vspace*{1cm}
\caption{\it Production of $K^*\bar{K}^*$ due to shake-out  of a
polarized \s\ pair from the proton wave function
in $\bar{p}p$ interaction from the initial $^3S_1$ state,
for (a) negative and (b) positive polarization of the \s\ pair.
The arrows show the directions of the spins of the nucleons and
quarks.}
\label{k*k*}
\end{figure}

The reason of these effects is easy to comprehend from Figs.
(\ref{kk}) and (\ref{k*k*}).

If the spins of the nucleon and antinucleon
are oriented in the same direction, as, e.g., in the $^3S_1$ initial
state, the shake-out of {\it negatively} polarized \s\
will form preferentially the {\it charged} pseudoscalar $K^+ K^-$
mesons from $s$ and $u$ quarks - which have opposite polarization - and
{\it neutral} vector
$K^{*0} \bar{K}^{*0}$ mesons,
from $s$ and $d$ quarks - which have the same polarization. The
corresponding quark diagrams are shown in Fig. \ref{kk}a and \ref{k*k*}a.
On the other hand,
if the \s\ quarks are polarized {\it positively}, i.e., along the
direction of the nucleon spin,
then $s$ and $u$ quarks will have the same polarization and
they will form preferentially the
{\it neutral} pseudoscalar $K$ mesons and {\it charged} vector
mesons as seen in Figs.~\ref{kk}b and \ref{k*k*}b, respectively.

It is important to note that these effects should be absent
for \an\ from the spin--singlet initial state
$^1S_0$.

This phenomenon has indeed been observed in bubble-chamber
experiments~\cite{Bar.65,Bal.65}, where it was
found that \an~ into
two neutral $K^*$ dominates over charged $K^*$ formation.
For instance, according to~\cite{Bar.65},
$Y(\bar{p}p \to K^{*0}\bar{K}^{*0}) = (30 \pm 7)\cdot10^{-4}$, whereas
$Y(\bar{p}p \to K^{*+}\bar{K}^{*-}) = (15 \pm 6)\cdot10^{-4}$.
However, a word of caution is needed: these data are quite old and
evaluation of the yield of two broad resonances in
presence of other open channels was done in a too simplified way.

However, the tendency has recently been confirmed
by the
Crystal Barrel collaboration~\cite{Abe.97} in measurements of the
channel $\bar{p}p \to K^0_L K^{\pm} \pi^{\mp} \pi^0$
in \an\ at rest. It was found that, for \an\ from the
$^3S_1$ state, the ratio between neutral and charged $K^*$ production
is
\begin{equation}
\frac{K^*(neutral) \bar{K}^*(neutral)}
{K^*(charged) \bar{K}^*(charged)} \approx 3
\end{equation}
We are not aware of other theoretical arguments that explain this
unexpected selection rule. On the other hand, the polarized-strangeness
model provides a natural explanation of
this effect, making essential use of the sign of the polarized pair.
In remarkable consistency with this hypothesis,
this effect is absent for \an\
from the $^1S_0$ initial state, again as it should be for the
shake--out of the polarized \s\ pair.

Till recent time it was believed \cite{Ams.98}, \cite{Rot.93}
that in the production of $K\bar{K}$ system there
is a suppression of the isospin I=0 amplitude on factor 5-10 in comparison
with the I=1 one. This hierarchy is nicely agrees with the prediction
of the polarized strangeness model\cite{Ell.99}. However it was shown
\cite{Kud.00} that
this conclusion was followed from the error in the data analysis and the
magnitude of the I=0 and I=1 amplitudes is approximately the same. \\
~\\

\section{Theoretical views on large
OZI violation processes}

        Let us consider how different theoretical models treat
concrete experimental facts of a large OZI violation discussed in Sect. 3.

\subsection{$\bar{p} p \to \phi\gamma$}

        The largest apparent OZI violation was observed in this
 channel (see (\ref{Rgamma})). However this reaction was measured
 for the spin-singlet $^1S_0$ initial state. It does not quite
 match the prediction of the polarized strangeness model that \f~
 should be produced mainly from the spin-triplet initial states.

        In \cite{Ell.95} it was argued that if really \f~
production from the spin-triplets dominates, then one would
expect that the ratio $\phi\gamma/\omega\gamma$ will
increase for \an ~from low pressure hydrogen gas, where the P-wave
\an~ is dominant.
Till now these measurements have not been performed and the
puzzle remains unsolved.

Recently A.Kotzinian \cite{Kot.00} brings attention to the fact that polarization
of the vacuum strange quarks by a valence quark may result in
formation of the \s~pair with $J^{PC}=0^{++}$ but spins of the strange
and antistrange quarks are oriented in opposite directions.
Indeed, in Sect.4.2 we discuss that the main idea of
\cite{Alb.95} for polarization of the strange sea is the assumption
that a proton valence quark interacts strongly with a vacuum
{\it antistrange} quark when the quantum numbers of the $u\bar{s}$
system are
pseudoscalars. So, the spins of the proton valence $u$-quark
and vacuum $\bar{s}$ quark should be oriented in  opposite
directions:

\begin{equation}
u~\stackrel{\uparrow}{\bullet}~
\stackrel{\circ}{\downarrow}~\bar{s}~
\label{s1}
\end{equation}
The quantum numbers of the \s~pair should have the vacuum quantum
numbers, i.e. the total spin $S=1$, the orbital angular momentum $L=1$
and the total angular momentum $J=0$. To provide that it was assumed
\cite{Alb.95} that the spin of the $s$-quark should follow the
direction of the spin of $\bar{s}$ quark:
\begin{equation}
u~\stackrel{\uparrow}{\bullet}~
\stackrel{\circ}{\downarrow}~\bar{s}~
\stackrel{\circ}{\downarrow}~s~
\label{s2}
\end{equation}

However the direction of the $s$ quark spin  may also be opposite to the
$\bar{s}$ ones, preserving
the total \s~ spin S=1 and the demand of $J^{PC}=0^{++}$.
Choosing an axis $z$ in the direction of the spin of the $u$
quark, there must be configuration with $S_z=0$:

\begin{equation}
u~\stackrel{\uparrow}{\bullet}~
\stackrel{\circ}{\downarrow}~\bar{s}~
\stackrel{\uparrow}{\circ}~s
\label{s3}
\end{equation}
along with the configuration of (\ref{s2}) with
 the projection on
$S_z=-1$.

 Then the \f~ production due to the rearrangement diagrams
will be modified. If both nucleon and antinucleon have \s~pairs with
$S_z=0$, then as in the case with $S_z=-1$, the rearrangement will
create \f~ preferentially from the spin-triplet states. However, if in the nucleon
there is the \s~pair with $S_z=-1$, and in the antinucleon the \s~ pair
is in $S_z=0$ state, then the rearrangement could produce the
\f~ also from the spin-singlet initial state.
Then the \f~production in \p~\an~ could be symbolically depicted as follows:
\begin{equation}
u~\stackrel{\uparrow}{\bullet}~
\stackrel{\circ}{\downarrow}~\bar{s}~
\stackrel{\circ}{\downarrow}~s +
\bar{u}~\stackrel{\bullet}{\downarrow}~
\stackrel{\circ}{\uparrow}~s~
\stackrel{\circ}{\downarrow}~\bar{s} \Longrightarrow
\stackrel{\circ}{\downarrow}~\bar{s}~
\stackrel{\circ}{\downarrow}~s   \label{ss}
\end{equation}
here only the valence $u$ and $\bar{u}$ quarks of proton and \ap~ are shown, the arrows indicate the
direction of quarks spin.

To what extent this possibility could describe the
violation of the OZI-rule in the $\phi\gamma$ channel depends on
the relative probability of $S_z=-1$ and $S_z=0$ components of \s~pair in
the proton wave function. We will come back to the discussion of this
question in the next section.

There are other explanations of
the $\phi\gamma$ paradox. The reaction
\begin{equation}
\bar{p} + p \to \phi + \gamma \label{pg}
\end{equation}
is quite specific - in this process
all valence quarks from the initial state should annihilate. So gluons are
dominating in the intermediate state. In such processes, as we discussed in
Sect. 2.5,  the applicability of the OZI rule for prediction of \f/\om ~ratio
is questionable.
It is interesting
that a similar strong OZI violation was seen in $\bar{p} p \to \phi\phi$
reaction (see, (\ref{fifi2})), where also all the valence quarks of the initial state
annihilate completely.

        Another possible explanation comes from the fact that the amplitude of the
 reaction (\ref{pg}) is connected with the amplitude of \f~photoproduction
 $\gamma + p \to \phi + p$. In \f~ photoproduction it is well known that the
 $\phi/\omega$ ratio
 does not follow the mixing angle prediction (\ref{rfi}).
 Thus in the diffractive photoproduction of $\phi$ mesons
\cite{Goo.80} a large value of the $\phi/\omega$ ratio was found:
\begin{equation}
  \frac{ \gamma A \to \phi \pi^+ \pi^- A}{ \gamma A \to \omega \pi^+ \pi^- A}=
(97\pm19)\cdot10^{-3} \label{gam}
\end{equation}
The photon
could interact strongly as a $\bar{q}q$ state and in the
photoproduction the contribution of
\s ~ pair  in the initial state is non--negligible. So the
diffractive photoproduction of \f~ is not described by the
disconnecting quark diagram and one could not expect validity of
the OZI rule prediction (\ref{rfi}) for this process as well as
for the reaction (\ref{pg}).

        Concrete calculations along these lines were performed in
\cite{Loc.94}, \cite{Gut.99}. The vector dominance model (VDM) was
applied. The reaction was treated in two steps: \an~ $\bar{p}p \to
\phi + V$ into \f~ and
some vector meson V,  and conversion of the produced vector meson into a real
photon via VDM. In \cite{Loc.94} it was claimed that it was
possible to reproduce successfully experimental branching ratio of
the $\phi \gamma$ channel using as an input branching ratios of
$\phi \rho$ and $\phi \omega$ channel and assuming destructive
interference between the amplitudes of these reactions. However
the analysis of \cite{Gut.99} confirms this conclusion only if the
phase space factor is chosen in a standard two-body form
 $f=k^{2l+1}$ . If the phase space factor is
taken from parametrization of Vandermeulen \cite{Van.88} as $f=k
\cdot exp(-A \sqrt{s- m^2_X})$, where $k$ is the final state c.m.
momentum, $\sqrt{s}$ is the total energy, $m_X$ is the sum of mass of the particles in the
final state and value of parameter $A=1.2~GeV^{-1}$ is
obtained from the fit of momentum dependence of the cross section of various
\an~channels, then the corresponding branching ratio drops down by factor 10 and turns out to be
$BR(\phi\gamma)= 1.5 \cdot 10^{-6}$. This is to be compared with
the experimental result $BR(\phi\gamma)= (2.0\pm0.4) \cdot 10^{-5}$ \cite{Ams.98}.
The question of the phase space factor for \an~reactions has been
discussed many times (see,e.g., \cite{Ams.98}, \cite{Rei.91} and references therein).
It was agreed that the Vandermeulen factor  better reflects the
many-channel nature of the \an~at rest, where a number of channels
 are open and the phase space
does not follow the simple two-body prescription.
 The authors of \cite{Gut.99} conclude that "large
observed branching ratio for $\phi \gamma$ remains unexplained in
the framework of VDM" and we would like to join this
statement.

\subsection{$\bar{p} p \to \phi\pi$}

        The experimental facts for this reaction seem to match perfectly with
the polarized strangeness model. The opulent \f~ production is
seen for \an~ from the spin-triplet $^3S_1$ wave
\begin{eqnarray}
 R_{\pi}(\phi/\omega, ~^3S_1) & = & (120\pm 12)\cdot 10^{-3}~,\\
 R_{\pi}(\phi/\omega, ~^1P_1) & < & 7.2\cdot 10^{-3} ~~~~~~~~~~~~~~
 \mbox{, with~95\%~CL}~
\end{eqnarray}
whereas the ratio from the spin-singlet $^1P_1$ initial state
is comparable with the mixing angle prediction (\ref{rfi}).

However, it was noted \cite{Rek.97} that the explanation \cite{Ell.95} of the
$\phi\pi$ reaction as a rearrangement process meets with a problem.
It was assumed \cite{Ell.95} that there is no spin-flip of
strange quarks due to
the rearrangement diagram, shown in Fig.\ref{fig:8}. However it is not
possible to meet this condition in $\bar{p}p\to \phi\pi$ reaction
assuming that both strange and antistrange quarks in nucleon are
{\it negatively} polarized, i.e. that the projection of the total
spin of \s~pair is $S_z=-1$.

Indeed, this reaction is going from the $^3S_1$
initial state, the orbital momentum $L=0$ and
 the projection on axis $z$ of total angular momentum
 J coincides with the projection of the total spin  $m_J=m_{S_i}=\pm 1,0$, where
$S_i$ is the total spin of nucleons in the initial state. Let us
assume that $m_{S_i}=m_J=+1$. Then the projection of the spin of the strange quarks
will be $m_s=m_{\bar{s}}=-1/2$ and the projection of the spin of \f~should be
$m_{S_{\phi}}=-1 $ assuming that there is no spin-flip of the
strange quarks during the \an.
The total angular momentum of the final state $J_f$ is the
sum $J_f=S_\phi + L_f$,
where  $S_\phi$ is the spin of \f~meson  and $L_f$ is the orbital
angular momentum between the \f~ and $\pi$-mesons. The P-parity
conservation dictates that $L_f=1$ for \an~from $^3S_1$.
Then it is clear that it is not possible to add $|S_\phi, m_{S_{\phi}}>= |1, -1> $
with $|L_f, m_{L_f}>= |1, m_{L_f}>$ to obtain $|J_f,m_{J_f}>=|1,+1>$ for any allowed values of $m_{L_f}$.
Schematically this situation is depicted as follows:
\begin{equation}
N~\stackrel{\uparrow}{\bullet}~
\stackrel{\circ}{\downarrow}~\bar{s}~
\stackrel{\circ}{\downarrow}~s +
\bar{N}~\stackrel{\uparrow}{\bullet}~
\stackrel{\circ}{\downarrow}~s~
\stackrel{\circ}{\downarrow}~\bar{s}
\neq
\phi~\stackrel{\uparrow}{\bullet}~ +
\pi~\bullet
 \label{rek}
\end{equation}

It is clear that to solve this problem one should introduce either
spin-flip of the $s$ quarks during the \an~ or give up the
assumption about the negative polarization of the strange quarks
in the nucleon. The spin-flip explanation is physically less
motivated and to avoid
this problem in \cite{Rek.97} it is assumed that the strange
quarks are polarized {\it positively} with respect to the nucleon
spin.

However, the suggestion of \cite{Kot.00} to take into account
the \s~pair
with vacuum quantum numbers and $S_z=0$ could easily solve the
problem. In that case the $\bar{p}p \to \phi \pi$ reaction
could be considered as the rearrangement of two \s~ pairs with
the $S_z=0$. And no spin-flip of strange quarks is needed to
provide correct orientation of the \f~spin, as it is seen from
the following schematic diagram:

\begin{equation}
N~\stackrel{\uparrow}{\bullet}~
\stackrel{\circ}{\downarrow}~\bar{s}~
\stackrel{\uparrow}{\circ}~s
+
\bar{N}~\stackrel{\uparrow}{\bullet}~
\stackrel{\circ}{\downarrow}~s~
\stackrel{\uparrow}{\circ}~\bar{s}
=
\phi~\stackrel{\uparrow}{\bullet}~ +
\pi~\bullet
 \label{rek2}
\end{equation}

In \cite{Loc.94,Loc.95,Lev.94,Gor.96} it has been suggested that the anomalously high yield of
the $\bar{p}p \to \phi\pi^0$
channel could be explained~
by rescattering diagrams with OZI-allowed transitions
in the intermediate
state, e.g.,  $\bar{p}p \to K^*\bar{K} \to \phi\pi^0$.
Calculations are capable~\cite{Loc.94,Lev.94,Gor.96}
to provide a reasonable agreement with the experimental data on
the $\phi\pi$ yield for annihilation from the S-wave.
However, what is not yet explained in this approach is the
strong dependence of the \f\ yield on the spin of the initial state.
Why the $\phi\pi$ yield from the spin-singlet state is 15 times
less than from the spin-triplet state is absolutely unclear in
these models.

In \cite{Zou.96} it was assumed that a possible reason may be
that the total decay width of the \p~ atom for the $^1P_1$ state
with isospin I=1
may be anomalously suppressed. This suppression was
predicted in some optical potential models as consequence of
the isospin-mixing in the protonium wave function. However this
suppression should be effective not only on the $\phi\pi$ but also on
the $\omega \pi$ channel and \cite{Zou.96} predicted that the
$\phi/\omega$ ratio for \an~from the P state may be as large as from
the spin-triplet S-state. That is  at variance with the
experimental data of (\ref{r3s})-(\ref{r1p}). These data show
that the ratio $\phi/\omega$ for \an~ from the P-wave is about 10
times less than the corresponding ratio for \an~from the S-state.
It is due to the absence of any suppression of the $\omega \pi$
channel from the P-state.

Moreover, the experimental results on
$K^*K$ branching ratios from the S- and P-states do not quite fit with
the predictions of the two-step model \cite{Loc.94,Loc.95,Lev.94,Gor.96}.
 If the anomalously high yield of the $\bar{p}p \to \phi\pi^0$
is due to rescattering diagrams with $K^*\bar{K}$ in the intermediate
state $\bar{p}p \to K^*\bar{K} \to \phi\pi^0$, then  to explain the
suppression of the
$\phi\pi$ yield for annihilation from the P-wave,
one should infer that it is due to unusually small frequency
of the $K^*\bar{K}$ amplitude from the $^1P_1$ channel \cite{Zou.96}.
\par
To verify this conclusion,
 the spin-parity analysis of
  annihilation frequencies of the $K^*\bar{K}$ final state at
 different target densities was performed \cite{Pra.98}. The results are shown in the
 table~\ref{Kstar}.

The yields correspond to the case of the best fit (Solution I), to demonstrate the
robustness of the results, the yields for
 another set of isobars (Solution II) are also shown.

  One can see that the $^1P_1$ fraction of the
 $K^* \bar{K}$ annihilation frequency is not negligible.
 It is comparable with the
 $^3S_1$ fraction and increases with the decrease of the target density.
This dependence is opposite to that of the $\phi\pi$ yield which decreases
with the target density.

        However, these results could not completely exclude the rescattering
mechanism of \cite{Loc.94,Loc.95,Lev.94,Gor.96}. The reason is due to the
impossibility of distinguishing between the isospin $I=0$ and $I=1$
components of the $K^*\bar{K}$ amplitude. In principle,
it may occur that the observed increase of the $K^*\bar{K}$
yield in the P-wave is due to the
$I=0$ part of the amplitude.
 Whereas the $I=1$ $K^*\bar{K}$ state,
 allowed for rescattering into
 $\phi\pi^0$, could be suppressed for some unknown reasons.
The final answer to this question should be given by the coupled channels
analysis of the $K^+K^-\pi^0$, $K^0 K^{\pm} \pi^{\mp}$
 and $\pi^+\pi^- \pi^0$
final states.

   It has been suggested \cite{Dov89}
that the enhancement of the $\phi$ meson production
in certain $\bar{N}N$ annihilation channels might be due to
resonances.
Specifically, if there existed a
vector ($J^{PC}=1^{--}$) $\phi\pi$ resonance  close to
the $\bar N N$ threshold,
it might be
possible to explain the selective enhancement of the  $\phi\pi$
yield in S-wave  annihilation, and the relative lack of
 $\phi$'s in the P-wave annihilation.
    The best candidate for such a state is
one
with mass $M=1480\pm40$ MeV,  width $\Gamma=130\pm60$ MeV and
quantum numbers $I=1,~J^{PC}=1^{--}$, which was observed
\cite{Bit.87} in the $\phi\pi^0$ mass spectrum
in  the reaction $\pi^- p \rightarrow K^+ K^- \pi^0 n$ at 32.5 GeV/c,
and dubbed the C-meson.

However,
this resonance cannot explain the OZI rule violation observed
in the $\phi\gamma$ channel, which is a final state with different quantum
numbers. The experimental
status of the C-meson is unconfirmed. Although some  experiments
have found  indications for its existence
(for a review, see \cite{Lan.88}), but others have not. It was not
seen in $pp$ central production \cite{Arm.92}, in \ap~\an~at rest
\cite{Ams.98} and in recent measurements of the E852 Collaboration
\cite{Ada.01}. The latter  investigated the same reaction $\pi^-p \to
K^+ K^- \pi^0 n$ at 18 GeV. The partial wave analysis shows that the intensity of
$\phi \pi$ wave is quite smooth in the $KK\pi$ mass interval of 1.2-1.5
GeV. Indication on some enhancement is seen around 1.6 GeV.
However the statistics is too scarce and the authors concluded that no resonances
was seen in the $\phi \pi$ system, at least, in the C-meson region.

The predicted \cite{Dov89} isoscalar partner of the C-meson which
should couple to the $\phi\eta$ channel  also was not observed, and no
deviation from the OZI rule has been detected in this mode.
Therefore the resonance interpretation of the
apparent violation of the OZI rule is not accepted due to the absence of the
corresponding states.

\subsection{$\bar{p} p \to f_2'(1525)\pi^0$}

The discovery of the OBELIX collaboration \cite{Pra.98} of the strong
OZI rule violation for the tensor \ten~meson was predicted in the
framework of polarized strangeness model \cite{Ell.95}. It was
predicted that the violation should occur just for \an~from the
P-wave and the experiment has confirmed that. It was found that

\begin{eqnarray}
 R(f_2'(1525)\pi^0/f_2(1270)\pi^0)
 & = &(47 \pm 14 ) \cdot 10^{-3} ~,~~~~\mbox{S-wave}   \\
 & = &(149 \pm 20 ) \cdot 10^{-3} ~,~~~\mbox{P-wave}
\end{eqnarray}

That should be compared with the OZI-rule prediction (\ref{rf2})
\begin{equation}
R=\frac{\sigma(A+B \rightarrow f_2'(1525) + X)}
{\sigma(A+B\rightarrow f_2(1270) + X)}  = 16\cdot10^{-3}
\end{equation}

The production of $f'_2$ in the
$\bar{p}p \to f'_2 \pi^0$ reaction
was calculated in the rescattering model assuming OZI-allowed
transitions to the $K^*K$ and $\rho\pi$ intermediate states
\cite{Lev.95}.
The obtained production rates of $f'_2$
are rather small, about $10^{-6}$. That is about two orders of the
magnitude less than the experimental values measured by the OBELIX
collaboration \cite{Pra.98}.

Therefore, it turns out impossible to accommodate the large
yield of the tensor
\s~ state just from the P-wave within the framework of the
rescattering model which used
the meson loops in the
intermediate state.

\subsection{$\bar{p} d\to \phi n$}

One of the largest violation of the OZI rule occurs in the
Pontecorvo reaction of $\bar{p}d$ \an.

\begin{equation}
R= Y(\bar{p} d\to \phi n)/Y(\bar{p} d\to\omega n) = (156 \pm 29)\cdot10^{-3}
\end{equation}

This fact had been predicted in the polarized strangeness model \cite{Ell.95}
a few years before
the corresponding experiment was started.

The other approach to treat the   Pontecorvo reactions was suggested
in \cite{Kon.89,Kon.98}. These reactions were considered
as
two-step processes.  First,
two mesons are created in the
$\bar p$ annihilation on a single nucleon of the deuteron and then
one of them is absorbed by the
spectator nucleon. In this approach, the OZI violation in
the Pontecorvo reaction $\bar{p} d \to \phi n$ is simply a reflection of its
violation in the elementary act  $\bar{p} p  \to  \phi \pi^0 $.

The model provides a possibility to account for the
large
ratio between the \f~ and \om~ production \cite{Kon.98}:

\begin{equation}
R_{th}=\frac{Y(\bar pd\rightarrow\phi n)}{Y(\bar pd\rightarrow\omega n)}=
(192\pm27)\cdot 10^{-3} \label{rth}
\end{equation}

However the two-step model is in
serious doubts after measurements by the Crystal Barrel collaboration
of the Pontecorvo reactions with the open strangeness \cite{Abe.99}:
\begin{eqnarray}
  \bar{p} + d \to & \Lambda + K^0 \label{lk}\\
\bar{p} + d \to & \Sigma^0 + K^0 \label{sk}
\end{eqnarray}

        It is found \cite{Abe.99}
that the yields of these reactions are practically
equal,\\
$R_{\Sigma,\Lambda}= Y(\Sigma K)/Y(\Lambda K) = 0.92\pm0.15$,
 in a sheer discrepancy with the two-step model prediction \cite{Kon.89}
that the $\Sigma$ production  should be about 100 times less than
the \la~
production. It was predicted \cite{Kon.98} that $R_{\Sigma,\Lambda}=0.012$.
This hierarchy appears naturally in the
two-step model due to the fact that the $\bar{K}N \to \Lambda X$ cross
section is larger than
the $\bar{K} N \to \Sigma X$ one.

        The measured yields of  the reactions (\ref{lk})-(\ref{sk}) are
also at least by a factor 10 over the two-step
model prediction \cite{Kon.89}.

        Therefore, experiments on Pontecorvo reactions clearly indicate
the opulent production of additional strangeness either in the form of
\f~mesons or of the $\Lambda K$ and $\Sigma K$ pairs.

\subsection{$\bar{p} p\to \phi \eta, \phi\rho, \phi \omega ...$}

One of the main puzzles in the complicated picture of \f~production
in \ap~\an~at rest is to understand why the increasing of the
\f~yield is observed only in some channels, like $\phi\gamma$ and
$\phi\pi$ whereas no deviation from the OZI predictions exists in
the other channels, like $\phi\eta, \phi\omega,\phi\rho$.

\begin{figure}[htb]
\begin{center}
\epsfig{file=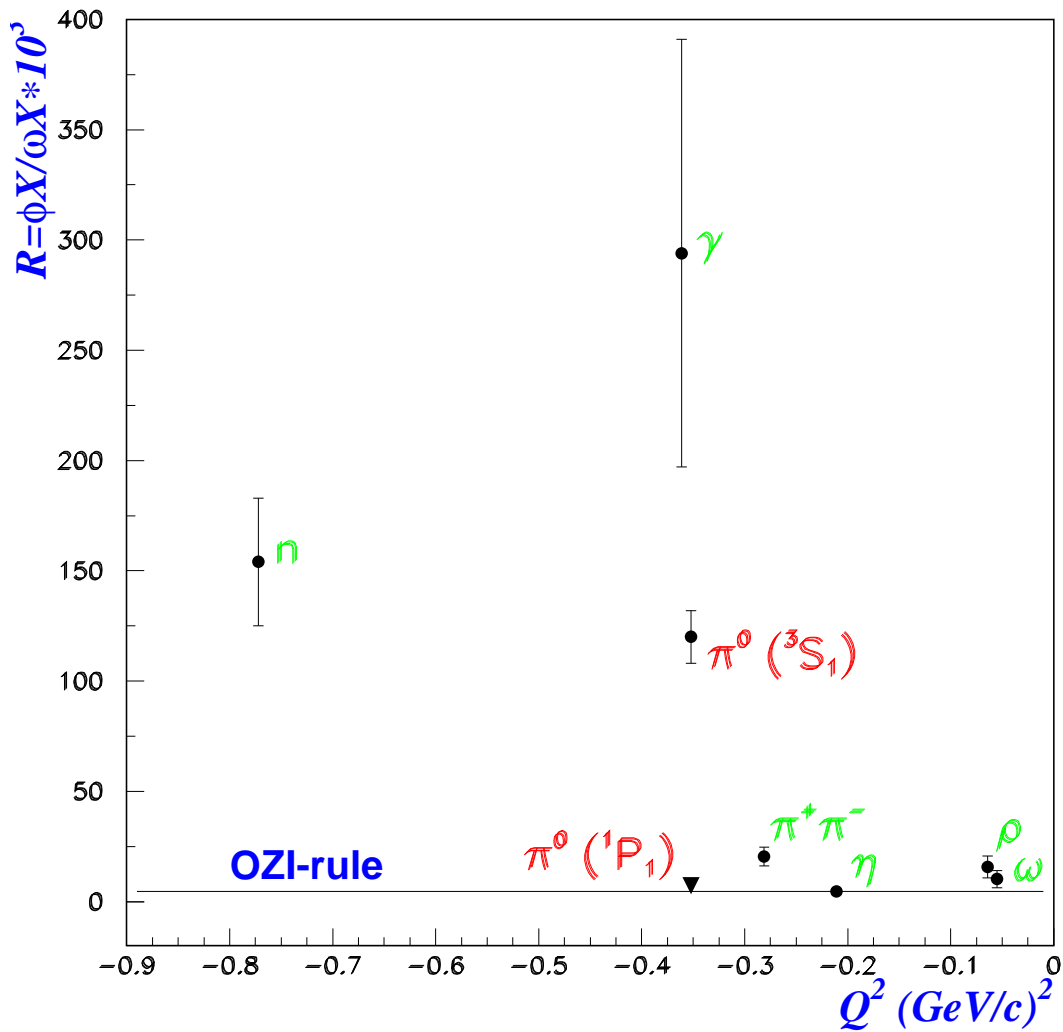,width=14.cm}
\caption{The ratio $R=\phi X/\omega X \cdot 10^3$ of yields for different reactions of $\bar pp \to \phi (\omega) X$
annihilation at rest as
a function of the momentum transfer to \f~.The solid line shows the prediction of the OZI rule (\ref{rfi}).}
\label{ozi}
\end{center}
\end{figure}

A possible key for solving this problem  is in the dependence on the momentum
transfer. In Fig. \ref{ozi} the
compilation of the data on the ratio
$R=(\phi X/\omega X) \cdot 10^3$ of yields for different reactions of $\bar pp \to \phi (\omega) X$
annihilation at rest is shown as a
function of the momentum transfer to \f~.
The solid line corresponds to the prediction of the OZI rule (\ref{rfi}).

One could see that the largest OZI-violation has
been observed for the reactions with the largest momentum transfer to
\f. That is Pontecorvo reaction $\bar{p}d\to \phi n$ and $\bar pp \to \phi\gamma, \phi \pi$
processes. The degree of the violation smoothly decreases with
mass of system $X$ created with the \f, i.e. with decreasing of
the momentum transfer. Thus for the $\phi\pi\pi$ final state with
light effective masses of the two-pions system around 300-400 MeV the
deviation from the OZI rule is significant. Whereas for the
$\phi\eta$ final state there is no problem with the OZI rule.

The polarized strangeness model explained this trend
 due to the
rearrangement
nature of the \f~production. The rearrangement
mechanism implies that two nucleons should
participate in the \f~production. This means a dependence on quantum numbers of both
nucleons as well as appearance of some minimal momentum transfer from
which this additional mechanism becomes important.

The rearrangement nature of additional \f~ production allows to
make some interesting prediction concerning \an ~ into the $\phi \eta$ final state.
As we discussed in Sect. 3.1.3 it was found \cite{Nom.98} that
the yield of the $\bar{p}p \to \phi \eta$ channel grows
with decreasing of the target density.
The branching ratio for \an ~ from the $^1P_1$ state turns out to be
by
10 times higher than that of the $^3S_1$ state:
\begin{eqnarray}
B(\bar{p}p \to\phi\eta, ^3S_1) & = &(0.76\pm 0.31)\cdot10^{-4} \\
B(\bar{p}p\to\phi\eta, ^1P_1) & = &(7.72\pm 1.65)\cdot10^{-4}
\end{eqnarray}

The polarized strangeness model suggests the following explanation of these facts:
the momentum transfer in the
$\phi \eta$ reaction is too small for the rearrangement diagrams
starting relevant. Increasing of the $\phi\eta$ yield in the P-wave is
not connected with proton intrinsic strangeness.
Therefore no OZI rule violation should be found neither for \an~ from the S-wave
nor from the P-wave. It means that the ratio $Y(\phi \eta)/Y(\omega \eta)$
should remain small in the P-wave. Therefore ten times increasing of the
$\omega \eta$ yield for \an~ from the P-wave is predicted.

Unfortunately, the kinematics of \ap~\an~ at rest restricts the
variation of the momentum transfer. It is important to study the
dependence of
the violation of the OZI rule on the  momentum transfer directly for \an~ in
flight.

\subsection{$p p \to p p \phi$}

       The production of \f~ in  nucleon-nucleon interaction will
provide a crucial test for the polarized intrinsic strangeness model
as an explanation of the strong OZI-violation seen in \ap~\an~at
rest. First, an enhancement of the \f~ production over OZI-prediction
should be seen. Second, a specific dependence of the \f~ production on the spin of the
initial NN state should
be observed. The \f~ should be produced mainly from the spin
triplet states.

In Sect. 3.2.1 we have discussed that the DISTO collaboration
\cite{Bal.98,Bal.00} indeed saw in the proton-proton collisions an enhancement of the
\f/\om~ratio on factor 10 over the OZI rule prediction. However, in this experiment the \f~ and
\om~ cross sections were evaluated at the same proton energy, it
means at different c.m. energies above the corresponding
thresholds. If one takes for the \om~ cross section the value, obtained
from the extrapolation of the energy dependence of all existing experimental
data, then the ratio \f/\om~ is still large, but it enhanced
over the OZI prediction by a factor 5 only \cite{Bal.00}. That
should be compared with enhancement factors 30-70 seen in some
reactions of \ap~\an~ at rest.

The theoretical analysis of the \f~and \om~ meson production was
done in \cite{Nak.98}-\cite{Tit.00}. The
diagrams of \f~production via mesonic current $\pi\rho \to \phi$
as well as \f~production via direct coupling with nucleon were
considered. It was found that the mesonic current dominates.
The contribution from the direct $\phi NN$ coupling is small. Precise size
of this contribution, evaluated in \cite{Nak.98},
 differs by a factor 4 for different sets of
parameters,
nevertheless the authors conclude that no violation of OZI-rule
is needed to invoke. However, a violation of OZI rule is needed in
the mesonic current, where the corresponding coupling
constants $\phi \rho \pi$ and $\omega \rho \pi$ turn out to be
connected in a different way than the OZI rule predicts. In other
words, it is not possible to explain both the $\omega$ and \f~ total and angular
cross sections assuming exact SU(3) relations between the coupling
constants $g_{\omega
\rho \pi}$ and $g_{\phi \rho \pi}$.

Recent experiment on the \om~production cross sections
\cite{Abd.01} in pp-collisions at 92 and 173 MeV excess energies
has
also confirmed inconsistencies arising in the theory \cite{Nak.99b} trying
to consider both the   \f~ and \om~mesons production within the same
approach. It turns out that the parameters of the mesonic current
chosen to fit the DISTO data on the \f~ cross section and angular distribution
overestimate the \om~ total cross sections and completely fail
to reproduce strong angular anisotropy of the \om~ angular
distribution.

However, the scarcity of the experimental data prevents from
any final conclusions. It is interesting to verify the polarized
strangeness model by using more clear tests.
In \cite{Ell.99} it was pointed out that
it is possible to verify the spin dependence of the \f\
production amplitude using unpolarized nucleons, comparing
\f\ production in $np$ and $pp$ collisions.
If \f\ is not produced from the spin--singlet states,
then the ratio of the $np$ and
$pp$ cross sections at threshold is
\begin{equation}
R_{\phi}=\frac{\sigma(np \to np\phi)}{\sigma(pp \to pp\phi)} =
\frac{1}{2} (1+ \frac{|f_0|^2}{|f_1|^2}) \approx \frac{1}{2}  \label{rfipp}
\end{equation}
In the framework of
the one-boson exchange model, i.e., without any assumption about the
nucleon's intrinsic strangeness,
this ratio was calculated~\cite{Tit.00} to be $R_{\phi}=5$.
Definitely, the experimental measurements of this ratio near threshold
could discriminate the predictions of these theoretical models.

\subsection{$\bar{p} p \to \phi \phi$}

As we discussed in Sect. 3.4 the opulent production
of the $\phi \phi$ system was seen in a number of experiments.
Thus the JETSET collaboration has seen an unusually
high apparent violation
of the OZI-rule in the
 $ \bar{p} + p \to \phi + \phi$
channel \cite{Eva.98}. The measured cross section of
this reaction
turns out to be two orders of the magnitude higher than the value
expected from the OZI-rule. The polarized strangeness model
predicted \cite{Ell.95}
that the $\phi \phi$ system
should be produced mainly from the initial
spin--triplet state.
Indeed, the data of the JETSET collaboration
\cite{Ber.96} have demonstrated that the initial spin--triplet state with $2^{++}$
dominates.

Moreover, it turns out that
the final states with the total spin $S$ of \f\f~ system  $S=2$ are
enhanced. This fact could be naturally explained in the polarized
strangeness model as consequence of the rearrangement of the \s~pairs from
proton and \ap.

        Of course, the polarized nucleon strangeness model is not
the only possible explanation of the facts.
In a model of
\cite{Mul.94} the \p~\an~ in \f\f~ was considered assuming
hyperon-antihyperon intermediate states, like $\Lambda
\bar{\Lambda}$, $\Sigma \bar{\Sigma}$. It turns out that the
calculated total cross section of \f\f~production agrees with
preliminary data of the JETSET collaboration  but missed
their final values \cite{Eva.98} by a factor 4.

The "simplest" explanation of the strong OZI violation in
the \f\f~ production is that the $2^{++}$ state dominance is a signal of a
tensor glueball  \cite{Bar.98},\cite{Etk.82}. So the coupling with
the tensor glueball of the proper mass increases the \f\f~ cross
section significantly above the prescription based on the mixing angle
value (\ref{fifi2}).

This possibility is remarkable in its general application to all
cases of the OZI violation. Everywhere the strange particle
production is enhanced, one could argue that it is due to the gluon
degrees of freedom. In hadron interactions at intermediate
energies there are not so many reactions where the effects of gluons
appeared explicitly, not hidden in the form of meson
and baryon exchanges. In this sense the OZI violation provides
the clear signal of importance of the gluon degrees of
freedom.

\section{Conclusions}

        This review has considered  the theoretical and
experimental situation with the fulfillment of the OZI rule - one
of the oldest phenomenological prescription of the hadron physics.
It turns out that the rule is working quite well in different
hadronic reactions at large energy interval. So, the general rule,
states that the processes describing by the disconnecting quark
lines is suppressed, is valid.

        However there are a number of experimental results
demonstrating a significant deviation from the predictions
(16)-(18),(\ref{fifi2}) based on
the OZI rule. Especially large violation (on factor 30-70) was
found in the \f~ production in the \ap~ \an~ at rest. The \f~meson
production in the low energy proton-proton interaction also
exceeds the OZI rule expectation by a factor 10-13.

An important
feature is the dependence of the degree of the OZI violation on
the spin of the initial state as well as on the sort of the meson,
created with the \f. It induces the idea that the observed violation
of the OZI rule is only apparent: the rule itself is valid but
the dynamics of the considered processes is not described by the
disconnecting quark diagrams. In particular, it is suggested
\cite{Ell.95}, \cite{Ell.99} that the physical reason for the
strong OZI violation in the strange particle production is
the polarization of the strange sea-quarks inside the nucleon.

        In Section 4 we collects the experimental and theoretical
results on the value of possible strange quark polarization
$\Delta s$. However this value was deduced from the {\it inclusive} DIS
data with some debatable assumptions.
In few experiments the sign of the charged hadrons in DIS was also measured
\cite{Ade.98}, \cite{Ack.99}.
The
semi-inclusive DIS data on charged hadrons  disagree with
the inclusive DIS results and claimed for quite small
polarization of the strange quarks. For instance, the HERMES result \cite{Ack.99} is
$\Delta s = -0.01\pm0.03\pm0.04$. However, an ad hoc assumption has been made in
this data analysis that the polarization of all sea quarks should
be equal. So it is up to a new generation of the semi-inclusive DIS
experiments, which could provide a good particle identification to
separate pions and kaons, to measure the $\Delta s$. At present
the best experimental possibilities exist for the HERMES (DESY) and
COMPASS (CERN) \cite{COMP}
experiments, where the corresponding programmes to measure the
polarization of different quark flavours are under way. Precise
determination of the strange and antistrange quarks polarized
structure functions
could be done in future  neutrino factories \cite{NF}.

 However even in the case of observation of non-zero polarization
of intrinsic nucleon strangeness it is absolutely non trivial that the
antiproton annihilation at rest or proton-proton interaction at threshold
will be affected by this effect. Therefore corresponding
measurements, confirming the OZI violation and investigating its
characteristics, are needed.

The polarization strangeness
model \cite{Ell.95}, \cite{Ell.99} could qualitatively explain
practically all experimental facts on strange particle production
in hadron interactions at low energies. However some caveats still
exist.

 Thus, the  strong OZI violation was seen in the  $\bar{p}p \to
\phi\gamma$ channel from the initial spin-singlet $^1S_0$ state. It
does
not fit with the polarized strangeness model postulate that \f~
should mostly be created from the spin-triplet initial states. To
cure the situation one should observe in this reaction even a
larger OZI violation from the initial spin-triplet $^3P_J$ states.
This experiment should be done.

Another problem appears to explain the spin transfer measurements
in the reaction $\bar{p} + \vec{p} \to \Lambda +
  \bar{\Lambda}$ \cite{Pas.00}.
The polarized strangeness model assumes an anticorrelation
  between spins of the proton and the s-quark. Then it is natural to predict
  a negative value for depolarization $D_{nn}$ measured in the
  \la~ production in
  polarized proton interactions $\vec{p}p \to \Lambda K^+ p$.
  The measurements  of DISTO collaboration \cite{Bal.99} indeed
 have confirmed this prediction. The same effect is expected for the
  depolarization of \la~ produced in \ap~ interactions with
  polarized protons $\bar{p} + \vec{p} \to \Lambda +
  \bar{\Lambda}$. However preliminary results of the PS 185
  experiment \cite{Pas.00} have shown that $D_{nn}$
  is quite small but the spin transfer to \al~$K_{nn}$ is unusually high
  and positive.

        There are versions of the polarized strangeness model
  where the spin of proton is indeed mainly transferred to
  \al~ rather than to \la~ \cite{Kot.00}. But in these modifications the
  $K_{nn}$ should be still negative. So, it is for the future to
  resolve this paradox.

\section{Acknowledgements}

        This review was ordered to us by late Prof. V.P.Dzhelepov.
        Unfortunately we are writing the text in quite a low pace during few
years. Finally it is over and we would like to pay a tribute
to V.P.Dzhelepov who inspired us to perform this work.

We are extremely grateful to  J.Ellis, S.B.Gerassimov, M.Karliner,
D.Kharzeev and  A.Kotzinian for the
numerous fruitful discussions.

\newpage
\section{Figures}

\newpage

\newpage

\newpage

\newpage

\vspace{1cm}

\vspace{1cm}

\newpage
\begin{table}[htb]
\caption{The ratio $R=\phi X/\omega X $ of the cross sections
for production of
$\phi$ and   $\omega$ - mesons in  $pp$, $\bar p p$ and   $\pi p$
interactions. $P_L$ is the momentum of the incoming particle.
 $Z$ is the parameter of the OZI--rule violation.
No corrections on the phase space volume difference were made
except the cases marked
$^{*)}$.}
\vspace*{0.5cm}
\begin{tabular}{crclll}
\hline
Initial & $P_L~~~$ & Final
& $R=\phi X/\omega X $
&$\left| Z\right|$ & Refs.\\
state & (GeV/c) & state $X$
& $~~~\cdot10^{3}$
&$(\%)$ & \\
\hline
$\pi^+ n$ & 1.54-2.6 & p & $21.0\pm11.0$
& $8\pm4$ &\cite{Dav.70},\cite{Dan.70} \\
$\pi^+ p$ & 3.54     & $\pi^+p$ & $19.0\pm11.0$
& $7\pm4$ & \cite{Abo.63} \\
$\pi^- p$ & 5-6      & n        & $3.5\pm1.0$
&  $0.5\pm0.8$ &\cite{Ayr.74} \\
$\pi^- p$ & 6        & n        & $3.2\pm0.4$
&  $0.8\pm0.4$ & \cite{Coh.77}\\
$\pi^- p$ & 10       & $\pi^-p$  & $6.0\pm3.0$
&  $1.3\pm2.0$ & \cite{Bal.77}\\
$\pi^- p$ & 19       & $2\pi^-\pi^+p$ & $5.0^{+5}_{-2}$
&  $0.6\pm2.5$&\cite{Woo.76} \\
$\pi^- p$ & 32.5     & n  & $2.9\pm0.9$
&  $1.1\pm0.8$ & \cite{Dor.95}\\
$\pi^- p$ & 360     & X          & $14.0\pm6.0$
&  $5\pm3$ & \cite{Agu.89}  \\
\hline
pp        & 10       & pp & $20.0\pm5.0$
&  $8\pm2$ & \cite{Bal.77}\\
pp        & 24       & pp & $26.5\pm18.8$
&  $10\pm6$ & \cite{Blo.75}  \\
pp        & 24       & $\pi^+\pi^-pp$ & $1.2\pm0.8$
&  $3\pm1$ & \cite{Blo.75} \\
pp        & 24       & pp m$\pi^+\pi^-$,
& $19.0\pm7.0$ & $7\pm3$ & \cite{Blo.75}\\
          &          &  m=0,1,2
&              &         &              \\
pp        & 70       &$ p X$
& $16.4\pm0.4$ &         & \cite{Gol.95}\\
pp        & 360     & X          & $4.0\pm5.0$
&  $0.1\pm4$ & \cite{Agu.91}  \\
\hline
$\bar p p$ & 0.7      &$\pi^+\pi^-$ & $19.0\pm5^{*)}$
&  $7\pm2$ &\cite{Coo.78} \\
$\bar p p$ & 0.7      & $\rho^0$      & $13.0\pm4^{*)}$
& $5\pm2$ &\cite{Coo.78} \\
$\bar p p$ & 1.2      &$\pi^+\pi^-$ & $11.0\pm^{+3}_{-4}$
&  $4\pm1$ &\cite{Don.77}  \\
$\bar p p$ & 2.3      &$\pi^+\pi^-$ & $17.5\pm3.4$
&  $7\pm1$ &\cite{Che.77}\\
$\bar p p$ & 3.6      & $\pi^+\pi^-$ & $9.0^{+4}_{-7}$
&  $3\pm3$ &\cite{Don.77}\\
\hline\\~\\
\end{tabular}
\label{tpp}
\end{table}

\begin{table}[htb]
\caption{The ratio $R=f_2'(1525) X/f_2(1270) X $
of the cross sections
for production of
$f_2(1270)$ and  $f_2'(1525) $ - mesons.
 $P_L$ is the momentum of the incoming particle.
In some cases,  marked by
$^{*)}$ , the cross section of $f_2(1270)$ was not separated from the
production of $a_2(1320)$ and the sum of these cross sections is given.}
\vspace*{0.5cm}
\begin{tabular}{ccccc} \hline
Initial state & $P_L$ , GeV/c & Final state X
 & $R(f_2'/f_2)\cdot 10^3$ & Ref. \\
\hline
$\pi^- p$ & $100.0$ & $n$ & $1.5\pm 0.7$ & \cite{ALDE86}   \\
$\pi^- p$ & $10.0$ & $n$ & $44\pm 10$ & \cite{COSTA80} \\
$\bar{p}p$ & $0 $ & $\pi^0$ & $26\pm 10$ & \cite{Ams.98} \\
$\bar{p}p$ & $2.32$ & $\pi^+\pi^-$ & $21\pm 10^{*)} $ & \cite{Chen} \\
$\bar{p}n$ & $2.3$ & $\pi^-$ & $<15$ & \cite{Gray} \\
$\bar{p}p$ & $1.5 - 2.0$ & $\pi^+\pi^-$ & $28\pm 9 ^{*)}$ & \cite{Vue}\\
$\bar{p}p$ & $7.02 - 7.57$ & $\pi^0$ & $64\pm 28 ^{*)}$ & \cite{Gag} \\
\hline
\end{tabular}
\label{tf2}
\end{table}

\begin{table}[tbh]
\caption{The ratio $R= \eta X/\eta' X $ for production of
$\eta$ and  $\eta'$ - mesons.
 $P_L$ is the momentum of the incoming particle.}
\vspace*{0.5cm}
\begin{tabular}{ccccc} \hline
Init. state & $P_L$ , GeV/c & X & $R(\eta/\eta')$ & Ref. \\ \hline
$\pi^- C$ & $38.0$ & $X$ & $2.0\pm 0.4$ & \cite{BANNIKOV89} \\
$\pi^- p$ & $8.45$ & $n$ & $2.5\pm0.2$ & \cite{STANTON80} \\
$\pi^- p$ & $15.0$ & $n$ & $1.7\pm 0.2$ & \cite{APEL79}\\
$\pi^- p$ & $20.2$ & $n$ & $2.0\pm 0.3$ & \cite{APEL79} \\
$\pi^- p$ & $25.0$ & $n$ & $1.9\pm 0.1$ & \cite{APEL79}  \\
$\pi^- p$ & $30.0$ & $n$ & $1.9\pm 0.1$ & \cite{APEL79} \\
$\pi^- p$ & $40.0$ & $n$ & $1.9\pm 0.1$ & \cite{APEL79}  \\
$\pi^- C$ & $39.1$ & $C^*$ & $1.7\pm 0.1$ & \cite{APOKIN86}  \\
$\pi^- p$ & $39.1$ & $n$ & $1.7\pm 0.2$ & \cite{APOKIN86} \\
$\pi^- p$ & $63.0$ & $n$ & $1.8\pm 0.6$ & \cite{DAUM80} \\
$\pi^- p$ & $300 $ & $n$ & $2.85\pm 1.06$ & \cite{Ald.89},\cite{Bar.98} \\
$\pi^+ p$ & $5.45$ & $\Delta^{++}$ & $4.5\pm 1.8$ & \cite{BLOODWORTH72}\\
$\pi^+ p$ & $16.0$ & $\Delta^{++}$ & $5.3\pm 3.4$ & \cite{HONECKER77}  \\
$pp$ & $450 $ & $pp$          & $2.21\pm 0.20$ & \cite{Bar.98}      \\
$\bar{p}p$ & $0$ & $\pi^0$ & $1.72\pm 0.21$ & \cite{Ams.98}  \\
$\bar{p}p$ & $0$ & $\omega$ & $1.94\pm 0.25$ & \cite{Ams.98}  \\
$\bar{p}p$ & $0$ & $\rho$ & $2.65\pm 0.79$ & \cite{Abe.9x},\cite{Urn.95} \\
$\bar{p}p$ & $0$ & $\pi^0\pi^0$ & $2.1\pm 0.5$ & \cite{Ams.9x},
\cite{Abe.B404}   \\
$\bar{p}p$ & $0$ & $\pi^+\pi^-$ & $2.2\pm 0.6$ & \cite{Abe.B411},
\cite{Urn.95} \\
$\bar{n}p$ & $x$ & $\pi^+$ & $1.59\pm 0.40$ & \cite{Fil.99}  \\
\hline
\end{tabular}
\label{teta}
\end{table}

\begin{table}
\caption[Reslt1]{
 The $\bar{p}p \to K^+K^-\pi^0$ and $\bar{p}p \to \phi\pi^0$
 annihilation frequencies (in units of 10$^{-4}$)
 for three densities of the hydrogen target \cite{Pra.98}.
 } \label{Reslt1}
\begin{center}
\begin{tabular}{lccc}
\hline\hline
 $Y\cdot10^{4}$  & LH$_2$ & NTP & 5~mbar \\
\hline\hline
$Y(\bar{p}p\rightarrow K^+K^-\pi^0)$ &
 23.7$\pm$1.6 & 30.3$\pm$2.0 & 31.5$\pm$2.2 \\
\hline
$Y(\bar{p}p\rightarrow \phi\pi^0)$ &
  4.88$\pm$0.32 & 2.47$\pm$0.21 & 0.92$\pm$0.10  \\
$Y(\bar{p}p\rightarrow \phi\pi^0,~^1P_1$) &
   &  & $<$0.1 with 95\%CL \\
\hline\hline
$Y(\bar{p}p\rightarrow \phi\pi^0$)
 & 3.3$\pm$1.5~\cite{Chi.88} & 1.9$\pm$0.5~\cite{Rei.91} &  \\
 other measurements
 &6.5$\pm$0.6~\cite{Ams.98} & 2.46$\pm$0.23~\cite{Abl.95b} &
 0.3$\pm$0.3~(LX)~\cite{Rei.91} \\
\hline\hline
\end{tabular}
\end{center}
\end{table}
\begin{table}
\caption[Kstar]{
  $K^*\bar{K}$ annihilation frequency (in units of 10$^{-4}$)
 at three densities of the hydrogen target \cite{Pra.98}.
 } \label{Kstar}
\begin{center}
\begin{tabular}{lccc}
\hline\hline
 Amplit. set ~~vs~~ $f(K^*\bar{K})\cdot f(K^*\bar{K}\rightarrow K^+K^-\pi^0)$
 & LH$_2$  & NTP &  LP (5~mbar)  \\
\hline\hline
 Solution I
                  & 5.20$\pm$0.89 & 7.01$\pm$0.84 &  7.45$\pm$0.95 \\
 $^3S_1$ fraction & 3.27$\pm$0.81 & 1.75$\pm$0.44 & 0.78$\pm$0.20  \\
 $^1P_1$ fraction & 0.89$\pm$0.16 & 3.44$\pm$0.53 & 4.52$\pm$0.70 \\
\hline
 Solution II
                  & 4.82$\pm$0.44 & 8.09$\pm$1.04 &  9.34$\pm$1.37 \\
 $^3S_1$ fraction & 2.47$\pm$0.24 & 1.29$\pm$0.13 & 0.57$\pm$0.08  \\
 $^1P_1$ fraction & 0.57$\pm$0.18 & 2.30$\pm$0.72 & 3.03$\pm$0.95 \\
\hline\hline\\
\end{tabular}
\end{center}
\end{table}
\end{document}